\documentclass[prd,twocolumn,amsmath,amssymb,showpacs,floatfix,
superscriptaddress,nofootinbib,preprintnumbers]{revtex4}

\usepackage{amsmath}
\usepackage{amsfonts}
\usepackage{graphicx}
\usepackage{dcolumn}
\usepackage{epsfig}
\usepackage{bm}
\usepackage{amsfonts}

\def\CC{{\cal C}}

\begin{document}

\title{Bipolar Harmonic encoding of CMB correlation patterns} \author{Nidhi
Joshi\footnote{niidhi.joshi@gmail.com}}
\author{S. Jhingan\footnote{sjhingan@iucaa.ernet.in}}
\affiliation{Centre for Theoretical Physics, Jamia Millia Islamia, New
Delhi-110025, India}

\author{Tarun Souradeep\footnote{tarun@iucaa.ernet.in}}
\affiliation{IUCAA, Post Bag 4, Ganeshkhind, Pune-411007, India}
\author{Amir Hajian\footnote{ahajian@princeton.edu}}
\affiliation{Department of Astrophysical Sciences,
Princeton University, Princeton, NJ 08544}

\begin{abstract}
Deviations from statistical isotropy can be modeled in various ways,
for instance, anisotropic cosmological models (Bianchi models),
compact topologies and presence of primordial magnetic field.
Signature of anisotropy manifests itself in CMB correlation
patterns. Here we explore the symmetries of the correlation function
and its implications on the observable measures constructed within the
Bipolar harmonic formalism for these variety of models. Different
quantifiers within the Bipolar harmonic representation are used to
distinguish between plausible models of breakdown of statistical
isotropy and as a spectroscopic tool for discriminating between
distinct cosmic topology.
\end{abstract}
\pacs{98.70.Vc, 98.80.Es}

\maketitle

\section{Introduction}
The fluctuations in Cosmic Microwave Background (CMB) contain an
amazing amount of information about our universe. Detailed
measurements of anisotropy in the CMB reveal global properties,
constituents and history of the universe. In standard cosmology, CMB
anisotropy is assumed to be statistically isotropic and
Gaussian. Gaussianity implies that the statistical properties of the
temperature field can be completely characterized in terms of its mean
$<\Delta T>=0$, and auto-correlation function $C(\hat n_1, \hat
n_2)=<\Delta T(\hat n_1)\Delta T(\hat n_2)>$, where $\hat n
=(\theta,\phi)$, is a unit vector on the sphere. The angular brackets
$<..>$ denote ensemble expectation values, i.e, averages above are for
all possible realizations of the field over a sphere. Since we have
one CMB sky, that is just one out of all possible realizations, the
ensemble expectation value $C(\hat n_1, \hat n_2)$ can be estimated in
terms of sky averages only to a limited extent, depending on
underlying symmetries in $C(\hat n_1, \hat n_2)$. Under the usual
assumption of \textit{Statistical Isotropy} (SI), implying essentially
Einstein's cosmological principle for cosmological perturbations, the
correlation function is invariant under rotations. It implies the
correlation function $C(\hat n_1, \hat n_2) = C(\hat n_1. \hat
n_2)\equiv C(\theta)$, can be readily estimated by averaging over all
pairs of sky directions separated by an angle $\theta$.

Spherical harmonics form a basis of the vector space of complex
functions on a sphere, making them a natural choice for expanding
the temperature anisotropy field,
\begin{equation}\label{eq:sp-har}
\Delta T(\hat n)=\sum_{lm}a_{lm}Y_{lm}(\hat n).
\end{equation}
Here $\Delta T$ is the temperature fluctuation around some average
temperature $T$. The complex quantities $a_{lm}$ are drawn from a
Gaussian distribution, related to the Gaussian temperature anisotropy
as
\begin{equation}\label{eq:alm}
a_{lm} = \int d\Omega_{\hat n} Y^*_{lm}({\hat n})\Delta T ({\hat n})\,.
\end{equation}
The condition for SI now takes the form of a diagonal covariance
matrix,
\begin{equation}\label{eq:SI}
<a_{l_1 m_1}a^{*}_{l_2 m_2}>=C_{l_1}\delta_{l_1 l_2}\delta_{m_1 m_2}
\,.
\end{equation}
Here $C_{l}$ is the well known angular power spectrum. In the SI case,
the angular power spectrum carries complete information about the
Gaussian field, and the statistical expectation values of the
temperature fluctuations are preserved under rotations in the sky.
This property of CMB has been under scrutiny since the release of the
first year of WMAP data. Tantalizing evidence for statistical isotropy
violation in the WMAP data using a variety of statistical measures has
also been claimed in recent
literature~\citep{preferreddirections,Eriksen:2003db,Hansen:2004vq}.
However, the origin of these `deviations' from SI remains to be
modeled adequately. These deviations could be either genuinely
cosmological, or statistical coincidence, or residual foreground
contamination, or, a systematic error in the experiment and the data
processing. Hence, it is important to carry out a systematic study of
SI violations using statistical measures within a unified,
mathematically complete, framework. Moreover, it is important to
develop several independent statistical measures to study SI
violations that can capture different aspects of  any measured
violation and provide hints toward its origin.

While testing a fundamental assumption, such as SI, is in itself a
justifiable end, there are also strong theoretical motivation to
hunt for SI violations in CMB on large scales. Topologically compact
spaces~\citep{G.Ellis,Lachieze,Gott,Cornish1,Levin,Linde,Souradeep}
and anisotropic cosmological models~\citep{Ellis,Barrow,Tuhin},
Cosmological magnetic fields generated during an early epoch of
inflation~\cite{Ratra} can also lead to violation of SI~\cite{Ruth},
are but a few examples.

This paper focuses on linking measures of SI violation to the
reduced symmetries of the underlying correlation
patterns\footnote{In this paper, we use the term `correlation
patterns' to interchangeably refer to SI violation} in the CMB map
or the correlation function. While we present illustrative examples
of the symmetries from various mechanisms of SI violation, this
paper does not concern itself with a study of specific mechanisms.
We define, within the framework of Bipolar harmonic representation
of CMB sky maps, a number of observables that can be used to
quantitatively test SI.  We present a study of the properties of
bipolar measures as one systematically reduces the rotational
symmetries of the CMB correlations, as is expected in different
theoretical scenarios. We recapitulate the bipolar harmonic
representation and the definitions of a set of measurable quantities
representing SI violation in section \ref{Bipfor}. In Section
\ref{sec:symm}, these observable measures are computed for different
levels of residual rotational symmetries of CMB correlations. This
provides a clear understanding of the underlying symmetries revealed
through the different bipolar measures. Section~\ref{bianchi} deals
with bipolar formalism measurables using Bianchi template.
Section~\ref{disc} summarizing conclusions and discussions is
followed by appendices where details of the calculations leading to
results are presented.

\section{Bipolar formalism and the observable measures}\label{Bipfor}

Any deviations from SI introduces \textit{off-diagonal} terms in the
covariance matrix Eq. (\ref{eq:SI}), thereby making $C_{l}$ an
inadequate quantity to characterize the statistical properties of
the temperature field~\cite{Pogosyan}. Under such a situation Bipolar spherical
harmonic expansion, proposed by Hajian and
Souradeep~\citep{AH-TS-03,AH-TS-04,AH-TS-05,AH-TS-06,AH-TS-NC,SB-AH-TS},
proves to be the most general representation of the two point
correlation function, where the angular power spectrum $C_{l}$ is a
subset of Bipolar spherical harmonic coefficients (BipoSH). Two
point correlation function of CMB anisotropies can be expanded as
\begin{equation}\label{eq:BPOSH}
C(\hat{n}_{1,}\hat{n}_{2}) = \sum_{l_{1},l_{2},L,M}
A_{l_{1}l_{2}}^{L M}\{Y_{l_{1}}(\hat{n}_{1})\otimes
Y_{l_{2}}(\hat{n}_{2})\}_{L M} \;,
\end{equation}
here $A_{l_{1}l_{2}}^{L M}$ are Bipolar Spherical Harmonic
coefficients (BipoSH), $|l_1 - l_2| \leq L \leq (l_1 + l_2)$,
$m_1+m_2=M$, and $\{Y_{l_{1}}(\hat{n}_{1})\otimes
Y_{l_{2}}(\hat{n}_{2})\}_{L M}$ are Bipolar spherical
harmonics~\cite{varsha}. Bipolar spherical harmonics form an
orthonormal basis on $\textbf{S}^{2} \times \textbf{S}^{2}$, with
transformation properties under rotations similar to spherical
harmonics. The tensor product in harmonic space can be explicitly
written using Clebsch-Gordan coefficients
$\CC_{l_{1}m_{1}l_{2}m_{2}}^{LM}$ as,
\begin{eqnarray}
\{Y_{l_{1}}(\hat{n}_{1})\otimes Y_{l_{2}}(\hat{n}_{2})\}_{L M} =\nonumber\qquad\\
\sum_{m_{1}m_{2}}\CC_{l_{1}m_{1}l_{2}m_{2}}^{LM}Y_{l_{1}m_{1}}(\hat{n}_{1})\;
Y_{l_{2}m_{2}}(\hat{n}_{2}) \; .
\end{eqnarray}

\subsection{Bipolar spherical harmonic coefficients - BipoSH}

BipoSH can be extracted by inverse transformation of Eq.
(\ref{eq:BPOSH}), i.e., multiplying both sides of this equation by
$\{Y_{l'_{1}}(\hat{n}_{1})\otimes
Y_{l'_{2}}(\hat{n}_{2})\}^{*}_{L'M'}$, and using orthonormality of
Bipolar spherical harmonics. Hence, given a real space correlation
pattern BipoSH coefficients can be found using
\[
A^{L M}_{l_1 l_2} = \int d\Omega_{\hat{n}_1}d\Omega_{\hat{n}_2}C
(\hat{n}_1, \hat{n}_2)\{ Y_{l_1} (\hat{n}_1)
\otimes Y_{l_2} (\hat{n}_2)\}^{*}_{LM} .
\]
Since $C(\hat{n}_{1,}\hat{n}_{2})$ is symmetric under the exchange
of $\hat n_1$ and $\hat n_2$, this gives rise to following symmetry
properties of BipoSH:
\begin{eqnarray}\label{eq:BipSym}
A^{LM}_{l_1 l_2}&=&(-1)^{l_1+l_2-L}A^{LM}_{l_2 l_1} \nonumber \\
A^{LM}_{ll}&=&A^{LM}_{ll}\delta_{L,2k}\, \quad . \qquad  \qquad
k=0,1,2,3,\ldots\,.
\end{eqnarray}
Hence, $A^{LM}_{ll}$ exists for \textit{even} $L$ and vanishes
otherwise.  It was shown in~\cite{AH-TS-03} that the Bipolar
Spherical Harmonic (BipoSH) coefficients $A^{LM}_{l_1 l_2}$ are a
linear combination of elements of the harmonic space covariance
matrix including the off-diagonal elements that encode SI violation,
\begin{eqnarray}
A^{LM}_{l_1 l_2}=\sum_{m_1 m_2}<a_{l_1 m_1}a^{*}_{l_2
m_2}>(-1)^{m_2}C^{LM}_{l_1 m_1 l_2 -m_2} .
\end{eqnarray}
When SI holds the covariance matrix  is diagonal, Eq. (\ref{eq:SI})
and Clebsch property (\ref{eq:trieq}), therefore
\begin{eqnarray}
A^{LM}_{l_1 l_2}=(-1)^{l_1}C_{l_1}(2l_1 + 1)^{1/2}\delta_{l_1
l_2}\delta_{L0}\delta_{M0},
\end{eqnarray}
implying that $A^{00}_{l l}$ contains all the information on the
diagonal harmonic space covariance matrix given by $C_l$.

The well known power spectrum $C_l$ thus forms a subspace of
BipoSH~\cite{AH-TS-04}. Under SI, the only non-zero Bipolar spherical
harmonic coefficient will be $A^{00}_{ll}$ (equivalent of $C_l$), all
the rest must  vanish. The violation of SI thus implies $A^{00}_{ll}$
are not sufficient to describe the field. Hence, BipoSH proves to be a
better tool to test SI, as non-zero $A^{L M}_{l_1 l_2}$, other then
$L=0$ and $M=0$, terms should confirm its violation.

It is impossible to measure all $A^{L M}_{l_1 l_2}$ from just one
CMB map because of cosmic variance. Thus we need to combine them in
different ways to diagnose different aspects of SI violations.

\subsection{Bipolar power spectrum- BiPS}

The \textit{Bipolar Power Spectrum} (BiPS) is a rotationally
invariant, quadratic measure that can be constructed out
of BipoSH coefficients~\cite{AH-TS-03}. BiPS involves averaging over
BipoSH that reduces cosmic variance in comparison to a single CMB
map, however this does not erase all the SI signatures. BiPS is
defined as
\begin{eqnarray}
\kappa_L=\sum_{l_1,l_2,M}|A^{LM}_{l_1 l_2}|^{2} .
\end{eqnarray}
For statistically isotropic models $\kappa_L=\kappa_0\delta_{L0}$,
i.e., $\kappa_L=0 \quad \forall \quad L>0$. Thus a breakdown of SI
will imply non-zero components of BiPS. In real space, $\kappa_L$
can be expressed as
\begin{eqnarray}\nonumber
\kappa_L & = &\left(\frac{2L+1}{8\pi}\right)^{2} \int
d\Omega_{\hat{\bf n}_1}\int d\Omega_{\hat{\bf n}_2} \\
  & & \left[\int dR \, \chi^{L}(R) \, C(R\hat n_1,R\hat
n_2) \right]^{2},
\end{eqnarray}
where $C(R\hat n_1, R\hat n_2)$ is the correlation function after
rotating the coordinate system through an angle $\omega\ (0\leq
\omega \leq \pi)$, about an axis  $\textbf{n}(\Theta,\Phi)$. $R\hat
n_1\ \mbox{and} \ R \hat n_2$ are the coordinates of the pixels
$\hat n_1$ and $\hat n_2$ in the rotated coordinate system. The
rotation axis $\textbf{n}$, is characterized by two parameters
$\Theta\ (0\leq\Theta\leq\pi)$, and $\Phi\ (0\leq\Phi\leq\ 2\pi)$.
$\chi^{L}$, is the trace of finite rotation matrix in
$LM$-representation called the \textit{characteristic function}, and
it is invariant under rotation of coordinate system,
\begin{eqnarray}
\chi^{L}(R)=\sum_{M}D^{L}_{MM}(R) .
\end{eqnarray}
Here $dR$ is the volume element of the three-dimensional rotation
group given by
\begin{eqnarray}
dR=4\sin^{2} \left(\frac{\omega}{2}\right) \; d\omega \;\sin\Theta
\; d\Theta \;d\Phi .
\end{eqnarray}
A simplified expression for BiPS in real space is
\begin{eqnarray}
\kappa_L & = & \frac{(2L+1)}{8\pi^{2}}\int d\Omega_{\hat{\bf
n}_1}\int d\Omega_{\hat{\bf n}_2} C(\hat n_1,\hat n_2) \nonumber
\\ & & \int dR \, \chi^{L}(R) \, C(R\hat n_1,R\hat n_2) \,.
\end{eqnarray}
For statistical isotropic model condition
$\kappa_L=\kappa_{0}\delta_{L0}$ can be recovered using
orthonormality of $\chi^{L}(R)$,
\[
\int^{\pi}_0 \chi^{L}(R) \chi^{L'}(R)\sin^2 \frac{\omega}{2} d\omega
= \frac{\pi}{2}\delta_{L L'} .
\]

The BiPS of CMB anisotropy computed from the maps measured by WMAP
are consistent with SI, rulings out its radical
violation~\cite{AH-TS-NC}. An advantage of BiPS is that its
rotational invariance allows for constraints to be placed on the
presence of specific forms of CMB correlation patterns independent
of the overall orientation in the sky.

\subsection{Reduced Bipolar coefficients- rBipoSH}

In order to extract information on the orientation of SI violation,
or to detect correlation patterns in a specific direction in the
sky, the \textit{Reduced Bipolar coefficients}~\cite{AH-TS-06},
obtained as
\begin{equation}
A_{LM}=\sum^{\infty}_{l_1=0}\sum^{L+l_1}_{l_2=|L-l_1|}A^{LM}_{l_1 l_2},
\end{equation}
provide another set of measures. The summation of BipoSH over
spherical wave-numbers $l_1$ and $l_2$, reduces the cosmic variance
rendering these measurable from the single CMB sky map available.

Note that the summation involves both the terms $A^{LM}_{l_1 l_2}$,
and $A^{LM}_{l_2 l_1}$, that are related via symmetry properties Eq.
(\ref{eq:BipSym}). Thus for any such combination where $l_1+l_2-L$
is odd, these two terms will cancel each other leaving no
contribution to the summation. The reduced Bipolar coefficients
$A_{LM}$, by definition have the following symmetry
\begin{equation}
A_{LM}=(-1)^{M} A^{*}_{L-M},
\end{equation}
which indicates $A_{L0}$ are always real. When SI condition is
valid, the ensemble average of $A_{LM}$ vanishes for all non-zero
values of $L$,
\begin{equation}
 <A_{LM}>=0 , \qquad \forall \quad L \neq 0.
\end{equation}
These $A_{LM}$ coefficients fluctuate about zero in any given CMB
anisotropy map. Therefore, a statistically significant deviation from
zero would confirm violation of SI. Unlike BiPS, reduced Bipolar
coefficients are sensitive to orientation, hence they can assign
directions to correlation patterns of the map.

\subsection{Bipolar map}

It is possible to visualize correlation patterns using the
\textit{Bipolar map} constructed from the reduced Bipolar
coefficients $A_{LM}$  as~\cite{AH-TS-06},
\begin{eqnarray}
\Theta (\hat n)=\sum_{LM}A_{LM}Y_{LM} (\hat n)\,.
\end{eqnarray}
The Bipolar map from $A_{LM}$ is computed in the same way as the
temperature anisotropy map from a given set of spherical harmonic
coefficients, $a_{lm}$. Bipolar map can also be represented in terms
of \textit{Tripolar Spherical Harmonics} of zero angular momentum
(see appendix \ref{app:Bipolar_map} for details),
\begin{eqnarray}\label{eq:tripolar}
\Theta(\hat n) & = & \sum_{L, l_1, l_2}\int d\Omega_{\hat
n_1}d\Omega_{\hat n_2}C(\hat n_1,\hat n_2)(-1)^{l_1+l_2} \sqrt{(2L+1)} \nonumber \\
& & \delta_{\lambda L}\{Y_{L}(\hat n)\otimes \{Y_{l_1}(\hat
n_1)\otimes Y_{l_2}(\hat n_2)\}_{\lambda}\}_{00}.
\end{eqnarray}
The tripolar spherical harmonics are expressed as~\cite{varsha}
\begin{eqnarray*}
&&\{Y_{l_1}(\hat n_1)\otimes\{Y_{l_2}(\hat n_2)\otimes Y_{l_3}(\hat
n_3)\}_{l_{23}}\}_{LM}= \\\nonumber && \sum C^{LM}_{l_1 m_1 l_{23}
m_{23}}C^{l_{23}m_{23}}_{l_2 m_2 l_3 m_3}Y_{l_1 m_1}(\hat n_1)Y_{l_2
m_2}(\hat n_2) Y_{l_3 m_3}(\hat n_3),
\end{eqnarray*}
where the summation is carried over $m_1, m_2, m_3,$ and $m_{23}$. The
transformations under rotations of tripolar spherical harmonics are
identical to spherical harmonics. In particular, the tripolar scalar
harmonics, which are invariant under rotations, can be expressed as
follows,
\begin{eqnarray*}
& & \{Y_{l_1}(\hat n_1)\otimes\{Y_{l_2}(\hat n_2)\otimes
Y_{l_3}(\hat n_3)\}_{\lambda}\}_{00} =
(-1)^{l_1+l_2+l_3}\delta_{\lambda \, l_1} \\ & & \sum_{m_1 m_2 m_3}
\begin{pmatrix}
l_1 & l_2 & l_3 \\
m_1 & m_2 & m_3 \\
\end{pmatrix}Y_{l_1  m_1}(\hat n_1)Y_{l_2  m_2}(\hat n_2)Y_{l_3  m_3}(\hat n_3) .
\end{eqnarray*}
Orthogonality and normalization relation is as follows,
\begin{eqnarray*}
\int\int\int d\Omega_{\hat n_1}d\Omega_{\hat n_2}d\Omega_{\hat
n_3}\{Y_{l_1}(\hat n_1)\otimes\{Y_{l_2}(\hat n_2)\otimes
Y_{l_3}(\hat n_3)\}_{\lambda}\}_{LM}\\  \{Y_{l'_1}(\hat
n_1)\otimes\{Y_{l'_2}(\hat n_2)\otimes Y_{l'_3}(\hat
n_3)\}_{\lambda'}\}^{*}_{L'M'} \\ = \delta_{l_1 l'_1}\delta_{l_2
l'_2}\delta_{l_3 l'_3}\delta_{\lambda \lambda'}\delta_{L
L'}\delta_{M M'} .
\end{eqnarray*}
From Eq. (\ref{eq:tripolar}) its evident that under SI the Bipolar
map is invariant under the rotations, since tripolar scalar
harmonics are rotationally invariant and $C(\hat n_1,\hat n_2) =
C(R\hat n_1,R\hat n_2)$. Hence, the map gets contribution only from
the monopole term $A_{00}$,
\begin{equation}
\Theta = \frac{1}{2}\sum_{l}(-1)^{l}\sqrt{\frac{(2l+1)}{\pi}}C_{l} .
\end{equation}
Also, if the temperature map is rotated by a element of rotation
group, $``R"$ then Bipolar map also rotates identically (see Appendix
\ref{app:rotation}). For example, if you rotate the temperature map
about the z-axis by some angle $``\alpha"$,
\[
\Delta T(R (\theta, \phi)
)=\sum_{lm}a_{lm}Y_{lm}(\theta,\phi-\alpha),
\]
the Bipolar map will also be rotated about z-axis through same angle
$``\alpha"$
\[
\Theta (R (\theta, \phi))=\sum_{LM}A_{LM}Y_{LM}(\theta,\phi-\alpha)
.
\]
However, the Wigner-D matrices in the two cases will be different
because of different $m$ (or $M$) values.

\section{Bipolar representation of CMB correlation
symmetries}\label{sec:symm}

The homogeneity and isotropy of cosmic microwave background points
to the Friedmann-Robertson-Walker(FRW) model of universe. Flat FRW
model adequately describes the observed local properties of the
universe, but the fact that universe with same local geometry can
admit different global topology has been appreciated since the
advent of post GR modern cosmology. This is because Einstein's
equations describe local properties of the spacetime and can only
constrain, but not determine, the global topological structure.

Symmetries of the space are preserved in the correlation function
and global topology modifies correlation function. The simply
connected (topologically trivial) hyperbolic 3-space ${\cal H}^3$,
and the flat Euclidean 3-space ${\cal E}^3$, are non-compact and
have infinite volume. There are numerous theoretical motivations,
however, to favor a spatially compact
universe~\citep{G.Ellis,Lachieze,Gott,Cornish1}. Compact topologies
(more, generally, multiply connected space) break the statistical
isotropy of CMB in characteristic patterns and induce a cutoff in
the power spectrum because of finite spatial
size~\cite{Tarun,Bond,Pogosyan,Weeks}. Theoretical possibilities
include compact Euclidean and Hyperbolic 3-spaces which require the
space to be multiply connected. The compact hyperbolic manifolds are
not globally homogeneous and they turn out to be not of much
interest for the class of symmetries considered under the scope of
this paper.

Simply connected universes are statistically isotropic, i.e. $C(\hat
n_1, \hat n_2)=C(\hat n_1.  \hat n_2)$. In contrast, all compact
universe models with Euclidean or hyperbolic geometry $C(\hat n_1,
\hat n_2)$ are statistically anisotropic. The isotropy of space is
broken in multi-connected models; this breaking of symmetry may be
apparent through the presence of some principal directions. In a
cylinder, for instance, which is compact in one dimension and
infinite in the other two, the metric tensor is exactly the same at
every point hence it preserves local homogeneity. However, it is not
globally isotropic and does not have the maximal symmetry. It is
noteworthy that globally anisotropic models do not contradict
observations, since the homogeneity of space and the local isotropy
can ensure the observed isotropy of the CMB, however can influence
the spectrum of density fluctuations. Multiply-connected models with
zero or negative curvature can be compact in some, or all their
dimensions. For instance a toroidal universe, despite its zero
spatial curvature, has a finite volume which may in principle be
measured. It contains a finite amount of matter. A cylindrical
universe (in the sense that the spatial sections are cylinders), on
the other hand, is noncompact in one dimension only and has an
infinite volume, although with a finite circumference in the
principal direction.

Homogeneity and isotropy are experimentally confirmed in the
observations of distribution of luminous red galaxies~\cite{hogg},
and the isotropy of CMB background~\cite{Bennet03, spergel}. Most of
the studies in CMB assume statistical isotropy of the universe (FRW
model). However, indications for a preferred direction in CMB, have
motivated the study of departures from statistical
isotropy~\cite{preferreddirections}. These deviations can arise from
non-trivial spatial
topologies~\citep{G.Ellis,Lachieze,Levin,Linde,Souradeep}, or
departures from the background FRW metric~\cite{Ellis, krasinski}.
Alternatively, statistical anisotropies might also arise from
coherent magnetic fields in the
universe~\citep{Ratra,Ruth,Rubenstein}. Anisotropic Cosmological
models have been considered in the past and they lead to
characteristic patterns in the CMB sky~\cite{Tuhin}. The Bianchi
template is an example of SI violation due to departure from
background FRW geometry. Here we will discuss the signature of
anisotropy due to existence of preferred axis (axes) on BipoSH. Such
SI violations can arise due to non-trivial topologies as well as
coherent magnetic fields.

Since Bipolar formalism is sensitive to structures and patterns in
the underlying two point correlation function, particularly the real
space correlations, it is a novel tool to characterize statistical
anisotropies
~\citep{AH-TS-03,AH-TS-04,AH-TS-05,AH-TS-06,AH-TS-NC,SB-AH-TS}.
Rotational symmetry about a preferred axis (say $\hat z$) is the
simplest way to break SI.

In general, the correlation function may be decomposed into
isotropic and anisotropic parts~\cite{Bond},
\begin{equation}
C(\hat{n}_{1,}\hat{n}_{2}) = C^{(I)}(\hat{n}_{1,}\hat{n}_{2}) +
C^{(A)}(\hat{n}_{1,}\hat{n}_{2}) .
\end{equation}
where
\begin{equation}
C^{ (I)}(\hat{n}_{1,}\hat{n}_{2}) = C(n_1 \cdot n_2)= \sum_{l}
\frac{2 l +1}{4\pi}~ C_l P_l(\hat{n}_{1} \cdot \hat{n}_{2})\, ,
\label{defisocorr}
\end{equation}
and the anisotropic part $ C^{(A)}$ is orthogonal to the Legendre
polynomials
\begin{equation}
\int d\Omega_{\hat n_1}\int d\Omega_{\hat n_2}\, C^{(A)}(\hat
n_1,\hat n_2)\, P_{l}(\hat n_1 \cdot \hat n_2) = 0\,.
\label{defanisocorr}
\end{equation}
This decomposition is useful in our study of the symmetries of the CMB
correlation patterns/structure that are explicit in real space.

\subsection{Statistical Isotropy (Rotational symmetry)}

Under SI, the correlation function is a function only of $\theta$,
the angle between the two directions, say, $\hat n_1$ and $\hat
n_2$. Hence, $C(\hat n_1 , \hat n_2) \equiv C(\hat n_1 \cdot \hat
n_2)=C(\theta)$, and the correlation function can be expanded in
terms of Legendre polynomials
\begin{equation}
C^{(I)}(\theta)=\frac{1}{4\pi}\sum_{l=2}^{\infty}(2l+1)C_{l}P_{l}(\cos\theta),
\end{equation}
where $C_{l}$ is the angular power spectrum. The summation starts from
$l=2$, since $l=0$ and $l=1$, respectively, monopole and dipole, are
usually subtracted out. For SI the angular power spectrum $C_{l}$
contains all the information.

In Bipolar representation, the condition of SI for various observables, described
in the previous section, can be summarized as
follows~\cite{AH-TS-03,AH-TS-06}:

\begin{itemize}
 \item BipoSH :
 $ A^{LM}_{l_1 l_2}=(-1)^{l_1}C_{l_1}(2l_1+1)^{1/2}\delta_{l_1l_2}\delta_{L0}\delta_{M0}, $

\item BiPS : $ \kappa_L=\kappa_{0}\delta_{L0}, $

\item rBipoSH : $A_{LM}=\sum_{l_1}(-1)^{l_1}C_{l_1}(2l_1+1)^{1/2}\delta_{L0}\delta_{M0}, $

\item Bipolar map: $ \Theta = \frac{1}{2}
\sum_{l}(-1)^{l}\sqrt{\frac{(2l+1)}{\pi}}C_{l} .
$

\end{itemize}
Therefore, to test a CMB map for statistical isotropy, one should
compute the BipoSH coefficients for the maps and look for non-zero
BipoSH coefficients. Cosmic variance calculated for BipoSH under
statistical isotropy is (see Appendix \ref{app:variance}) ,
\begin{eqnarray}
\sigma^{2}_{SI}(\tilde {A}^{LM}_{l_1 l_2})= C_{l_1}C_{l_2}[1+(-1)^{L}\delta_{l_1 l_2}]
\end{eqnarray}
for rBipoSH is,
\begin{eqnarray}
\sigma^{2}_{SI}(\tilde {A}_{LM})= \sum_{l_1 l_2}C_{l_1}C_{l_2}[1+(-1)^{l_1+l_2-L}]
\end{eqnarray}
and for BiPS~\cite{AH-TS-03, Amirthesis},
\begin{widetext}
\begin{eqnarray}
\sigma^{2}_{SI}(\kappa_L)= \sum_{l:2l\geq
L}4C^{4}_{l}[2\frac{(2L+1)^{2}}{2l+1}+(-1)^{L}(2L+1)+(1+2(-1)^{L})F^{L}_{ll}]+\sum_{l_1}\sum^{L+l_1}_{l_2=|L-l_1|}4C^{2}_{l_1}C^{2}_{l_2}[(2L+1)+F^{L}_{l_1
l_2}]\nonumber
\\+8\sum_{l_1}\frac{(2L+1)^{2}}{2l_1+1}C^{2}_{l_1}[\sum^{L+l_1}_{l_2=|L-l_1|}C_{l_2}]^{2}+16(-1)^{L}\sum_{l:2l\geq
L}\frac{(2L+1)^{2}}{2l_1+1}\sum^{L+l_1}_{l_2=|L-l_1|}C^{3}_{l_1}C_{l_2}
\end{eqnarray}
\end{widetext}
where
\begin{eqnarray}
F_{l_1l_3}^{L} &=& \!\!\!\!
\sum_{m_1m_2=-l_1}^{l_1}\,\sum_{m_3m_4=-l_3}^{l_3}
\sum_{M,M'=-L}^{L} C^{L M}_{l_{1}-m_{1}l_{3}-m_{3}}C^{L
M}_{l_{1}m_{2}l_{3}m_{4}}\nonumber \\&& \times C^{L
M'}_{l_{3}m_{4}l_{1}m_{ 1}}C^{L M'}_{l_{3}-m_{3}l_{1}-m_{2}}
\end{eqnarray}
and $L$ is even. Statistically significant deviations from zero
would mean violation of statistical isotropy.

\subsection{Cylindrical  symmetry}
The correlation function must satisfy the symmetries of the
underlying theory. In Friedman models the symmetry group is SO(3),
hence the correlation function is invariant under rotations; any
breakdown of SI will reduce this symmetry group. The simplest way to
break SI is to introduce a favored direction in the sky, in such a
case the reduced symmetry group is SO(2) or cylindrical symmetry.
Assuming the favored axis to be z-axis, the rotational symmetry
about z-axis for any arbitrary $\Delta \phi$ will require,
\begin{eqnarray*}
C^{(A)}(\theta_1, \phi_1, \theta_2 ,\phi_2)=C^{(A)}(\theta_1
,{\phi_1}+\Delta\phi, \theta_2, {\phi_2}+\Delta\phi).
\end{eqnarray*}
where $n_1 \equiv (\theta_1,\phi_1)$ and $n_2 \equiv  (\theta_2,
\phi_2)$. The most general form of the correlation function in such
a case is (see Appendix \ref{app:zerofold})
\begin{eqnarray}\label{eq:zerofold}
C^{(A)}(\theta_1,\phi_1,\theta_2,\phi_2)=\sum_{m}\textit{f}_m(\theta_1,\theta_2)\cos
m(\phi_1-\phi_2).
\end{eqnarray}
Further, if the correlation function is invariant under the
reflection, i.e., looking at a correlation pattern in the sky one
cannot distinguish whether we are looking up or down the preferred
direction, then
\begin{eqnarray}
C(\pi-\theta_1,\phi_1,\pi-\theta_2,\phi_2)=C(\theta_1,\phi_1,\theta_2,\phi_2),
\end{eqnarray}
which leads to a condition that $l_1+l_2$ is even. BipoSH in such a
case would be, or equivalently the covariance matrix will be
~\cite{Ferreira:1997},
\begin{equation}
<a_{l_1 m_1}a^{*}_{l_2 m_2}>=\delta_{m_1 m_2}C^{l_1 l_2}_{|m_1|},
\end{equation}
where diagonal terms $C^l_{m}$ of $C^{l_1 l_2}_{|m_1|}$ are called
the cylindrical power spectrum, and $|m|>0$ modes are the allowed
frequencies for scale $l$. There may be correlations between various
scales called connectivity of fluctuations.  The expression for
$C^{l_1 l_2}_{|m_1|}$ in terms of the correlation function is

\begin{eqnarray*}
C^{l_1
l_2}_{|m_1|}=\frac{1}{8\pi}
\sqrt{\frac{(2l_1+1)(2l_2+1)(l_1-m_1)!(l_2-m_1)!}{(l_1+m_1)!(l_2+m_1)!}}
\times \nonumber\\
\int^{\pi}_{0}P_{l_1 m_1}(\cos\theta_1)P_{l_2
m_1}(\cos\theta_2)f_{m_1}(\theta_1,\theta_2)d(\cos\theta_1)d(\cos\theta_2)\,.
\end{eqnarray*}
Using Eq. (\ref{eq:bipocoeff}) in the appendix, we obtain
\begin{eqnarray*}
A^{LM}_{l_1 l_2}=[1+(-1)^{l_1+l_2-L}]\sum_{m}(-1)^{m}C^{l_1
l_2}_{|m|}C^{LM}_{l_1 m l_2 -m}\delta_{M0} .
\end{eqnarray*}

When $m$ is even, the functions $P^{m}_{l}$ will be odd or even
functions of their arguments, depending on whether $l$ is odd or
even respectively. Similarly, for the odd $m$'s. In both the cases
when only one of $l_1$ and $l_2$ is odd, the integral vanishes.
Therefore we have to consider cases when both of them are either odd
or even. In such a case $l_1 + l_2$ is even and hence $A^{LM}_{l_1
l_2}$ vanishes for $L=$ odd,
\begin{equation}
A^{LM}_{l_1 l_2}=A^{LM}_{l_1 l_2}\delta_{L,2k}\delta_{M0},\ \ {\rm where}\ k=0,1,2,3,....
\end{equation}

\begin{eqnarray}
A^{LM}_{l_1 l_2} =  A^{LM}_{l_1
l_2}\delta_{L0}\delta_{M0}\delta_{l_1 l_2}+A^{LM}_{l_1
l_2}\delta_{M0} .
\end{eqnarray}
Therefore, the BipoSH present under cylindrical  symmetry are
$A^{00}_{ll}$  and  $A^{L0}_{l_1 l_2}$ with even $L$. Using symmetry
property of BipoSH (\ref{eq:BipSym}), under cylindrical symmetry we
have $A^{LM}_{l_1 l_2}=A^{LM}_{l_2 l_1}$, i.e., the BipoSH are
symmetric under the exchange of $l_1$ and $l_2$. There is another
possibility that $a_{lm}$'s have a gaussian distribution with
different variance for each $m$ mode corresponding to a particular
$l$. This implies breakdown of SI, as power in each $m$ mode is
different, $C^{l_1 l_2}_{|m_1|}=\delta_{l_1 l_2}C^{l_1}_{|m_1|}$,
and the corresponding Bipolar coefficients are,
\begin{eqnarray}
A^{LM}_{l_1 l_2}=A^{LM}_{l_1 l_2}\delta_{L0}\delta_{M0}\delta_{l_1
l_2}+A^{LM}_{l_1 l_2}\delta_{M0}\delta_{l_1 l_2} .
\end{eqnarray}
In such a case non-zero BipoSH are $A^{00}_{ll}$ and $A^{L0}_{ll}$,
where multipole moment is even and $(L\geq2)$. Furthermore, it may
happen that a given model displays the degeneracy
$C^{l_1}_{|m_1|}=C_{l_1}$ and the rotational symmetry SO(3) of
covariance matrix is restored.  The rBipoSH for cylindrical symmetry
are,
\begin{equation}
A_{LM}=\sum_{l_1 l_2}A^{LM}_{l_1 l_2}=A_{LM}\delta_{L,2k}\delta_{M0}
\ \mbox{where} \ k=0,1,2,3,....
\end{equation}
Hence the Bipolar map for such a symmetry will be,
\begin{eqnarray}
\Theta(\hat n)&=&\sum_{LM}A_{LM}Y_{LM}(\hat
n)=\sum_{L}A_{L0}\delta_{L,2k}Y_{L0}(\hat n) \nonumber \\ &
=&\sum_{L}\sqrt{\frac{2L+1}{4\pi}}A_{L0}\delta_{L,2k}P_{L}(\cos\theta)
\end{eqnarray}
Thus the map here looks like a sphere which is divided into latitude
bands, or zones, without any longitudinal variation.

A realistic example of cylindrical symmetry is a primordial
homogeneous magnetic field which breaks statistical isotropy by
inducing a preferred direction (${\bf e}$). Therefore, the correlation
function between two points ($ {\bf n}$ and ${\bf n}'$) depends not
only on the angular separation between two points (${\bf n}.{\bf n}'$)
but also on their orientation with respect to the magnetic field. This
dependence of correlation function on angles between ${\bf n}$ and
${\bf e}$ (as well as ${\bf n}'$ and ${\bf e}$) leads to correlation
between $l$ and $l \pm 2 $ modes ~\cite{Ruth}. The vector nature of
the magnetic field induces off-diagonal correlations\cite{Ruth}
\begin{equation}
 D_{l}(m)=<a^{*}_{l-1 m} a_{l+1 m}> \equiv <a^{*}_{l+1 m}a_{l-1 m}>
 \, .
\end{equation}
Here $D_{l}$ is the power spectrum of the off-diagonal elements of
the covariance matrix, and the correlation function shows up as,
\begin{eqnarray}
<a_{l_1 m_1}a^{*}_{l_2 m_2}> &=& \delta_{m_1 m_2} \delta_{l_1
l_2}C_{l_1} +
\\ \nonumber && \delta_{m_1 m_2} (\delta_{l_1+1,l_2-1}+\delta_{l_1-1,l_2+1})D_{l_1} .
\end{eqnarray}
The BipoSH corresponding to this covariance matrix are
\cite{Amirthesis},
\begin{eqnarray}
A^{LM}_{l_1 l_2}&=&(-1)^{l_1}(2l_1+1)^{1/2} C_{l_1}\delta_{l_1
l_2}\delta_{L0}\delta_{M0} \\&+&D_{l_1}\delta_{l_1 l_2\pm 2}
\delta_{M0}\sum_{m}(-1)^{m}C^{L0}_{l_1 m l_2 -m}.\nonumber
\end{eqnarray}
The non-zero BipoSH in this case are $A^{00}_{ll}\ and \ A^{L0}_{l_1
l_1 \pm 2}$.

The reduced Bipolar spherical harmonic coefficients (rBipoSH) for
this case are
\begin{eqnarray}\label{eq:rbimag}
A_{LM}&=&\sum_{l}\delta_{M0}\delta_{L0}(2l+1)^{1/2}C_{l} \nonumber
\\& &  + 2\sum_{lm}(-1)^{m}D_{l}C^{L0}_{l-1 m l+1 m}\delta_{M0} .
\end{eqnarray}
These coefficients are non-zero only for $l_1+l_2-L = $ even, thus
$L$ can take only even values. Finally, the Bipolar map is,
\begin{eqnarray}
\Theta(\hat n)&=& \sum_{L}A_{L0}\delta_{L,2k}Y_{L0}(\hat n),\qquad
k=0,1,2,3 \nonumber \\
& = &
\frac{1}{2\sqrt{\pi}}A_{00}+\sum_{L}A_{L0}\delta_{L,2a}P_L(\cos\theta),\quad
a=1,2..\nonumber
\end{eqnarray}
where $A_{00}$ and $A_{L0}$ are given by Eq.(\ref{eq:rbimag}).

\subsection{n-fold discrete  Cylindrical symmetry}

\begin{figure}[ht]
\begin{center}
\mbox{\epsfig{figure=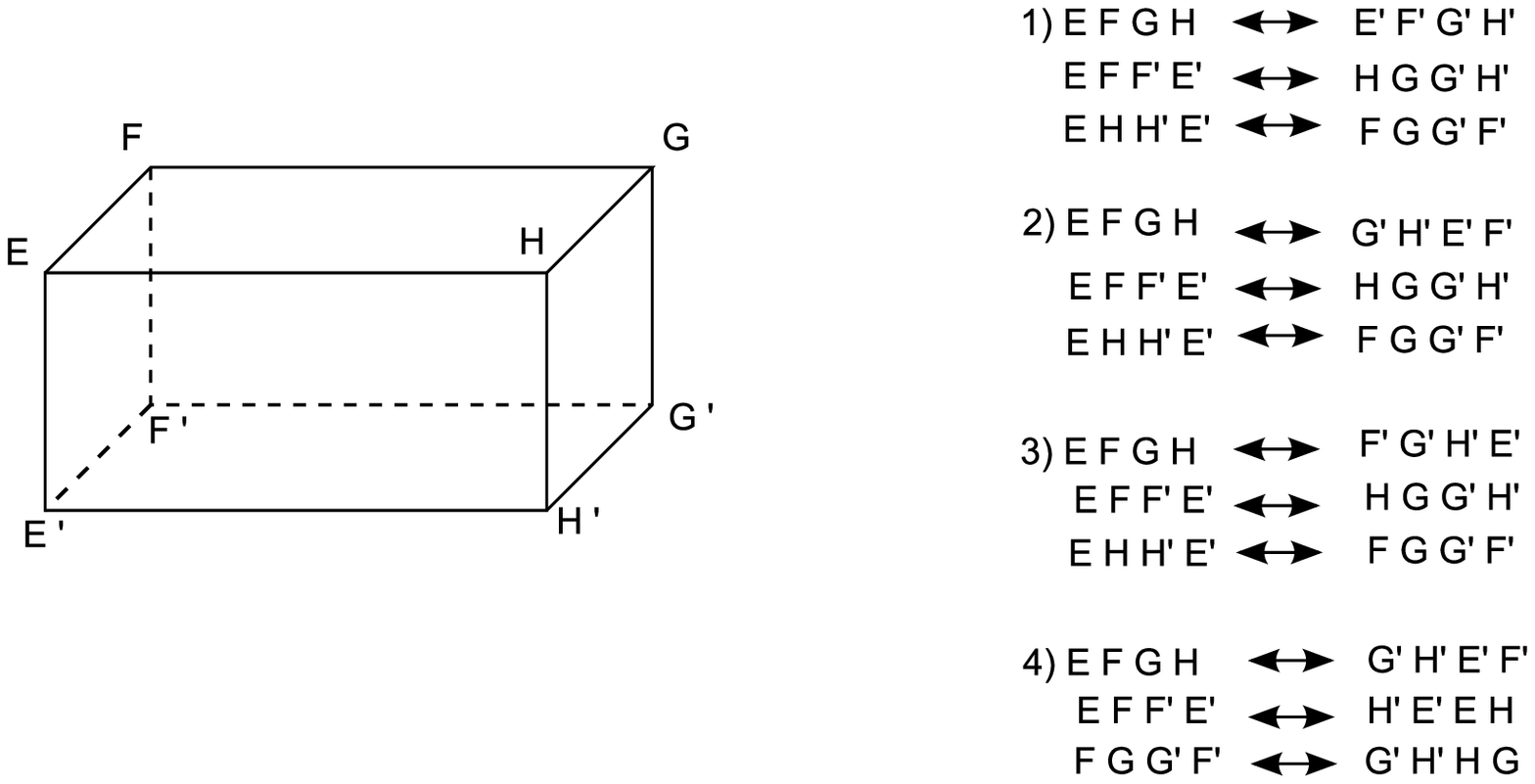,width=8.cm,angle=0}} \caption{ {\it
The locally Euclidean, closed, oriented 3-spaces.  }} \label{topo1}
\end{center}
\end{figure}

Violation of SI also manifests itself in compact universes with flat
universal cover, which exhibits a n-fold rotational symmetry about
an axis. There are six possible compact models of the universe
having a flat universal cover (UC)~\cite{Lachieze}. These are
visualized by identifying opposite sides of the fundamental
polyhedra. The fundamental polyhedron (FP) may be a parallelepiped.
The possible identifications then are (figure \ref{topo1})
\begin{itemize}
\item[1)] - opposite faces by translations.
\item[2)] - opposite faces, one pair being rotated by angle $\pi$.
\item[3)] - opposite faces, one pair being rotated by $\pi / 2$.
\item[4)] - opposite faces, all three pairs being rotated by $\pi$.
\end{itemize}
The fundamental polyhedron can also be the interior of an hexagonal
prism, with two possible identifications (figure \ref{topo2}) :
\begin{figure}[ht]
\begin{center}
\mbox{\epsfig{figure=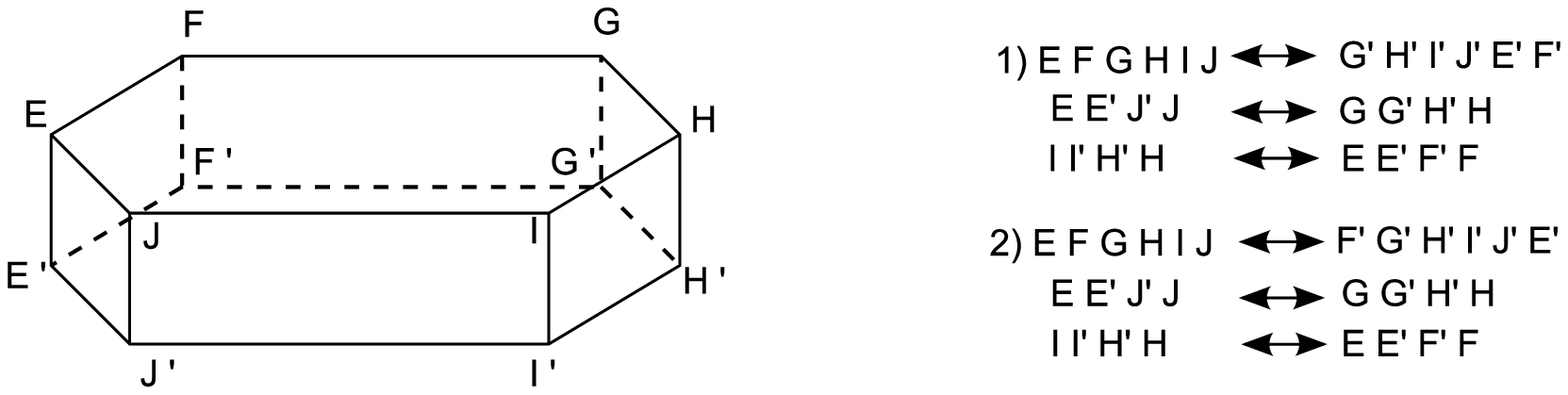,width=8.cm,angle=0}} \caption{ {\it
The locally Euclidean, closed, oriented 3-spaces.  }} \label{topo2}
\end{center}
\end{figure}
\begin{itemize}
\item[1)] - opposite faces, the top face being rotated by an angle $2 \pi / 3$
with respect to the bottom face.
\item[2)] - opposite faces, the top face being rotated by an angle $\pi / 3$
with respect to the bottom face.
\end{itemize}
Correlation function having a n-fold rotational symmetry about
z-axis can be written as
\begin{equation}\label{eq:nfoldsymm}
C^{(A)} (\theta_1,\phi_1,\theta_2,\phi_2) =
C^{(A)}(\theta_1,\phi_1+\frac{2\pi}{n},\theta_2,\phi_2+\frac{2\pi}{n}).
\end{equation}
This symmetry enforces $m_1+m_2=nk$, where $n$ can be odd or even,
depending upon the symmetry of the compact universe and
$k=0,1,2,3...$ . Thus the general form of correlation function is
(see Appendix \ref{app:nfold}),
\begin{eqnarray}
&&C^{(A)}(\theta_1,\phi_1,\theta_2,\phi_2) =
\sum_{m_1,m_2}f_{m_1,m_2}(\theta_1,\theta_2) \times \nonumber \\ & &
e^{i(m_1\phi_1+m_2\phi_2)}\delta_{m_1+m_2,nk},\ \ \ k=0,\pm 1,\pm
2...
\end{eqnarray}
Corresponding Bipolar spherical harmonic coefficients under the
symmetry eq.(\ref{eq:nfoldsymm}) are,
\begin{widetext}
\begin{eqnarray}
A^{LM}_{l_1 l_2}=\frac{1}{4\pi}\sum_{m_1 m_2}
\sqrt{\frac{(2l_1+1)(2l_2+1)(l_1-m_1)!(l_2-m_2)!}{(l_1+m_1)!(l_2+m_2)!}}
\delta_{m_1+m_2, nk}C^{LM}_{l_1 m_1 l_2 m_2} \int^{2\pi}_{0}\int^{2\pi}_{0}\int^{\pi}_{0}\int^{\pi}_{0}\times\nonumber\\C(\theta_1,\phi_1,\theta_2,\phi_2)
e^{-i(m_1\phi_1+m_2\phi_2)}P_{l_1 m_1}(\cos\theta_1)P_{l_2 m_2}(\cos\theta_2)
d(\cos\theta_1)d(\cos\theta_2)d\phi_1 d\phi_2.
\end{eqnarray}
\end{widetext}
All possible Euclidean models of compact universe exhibit reflection
symmetry about the xy-plane. The correlation function under
reflection symmetry is,
\begin{eqnarray}
C(\theta_1,\phi_1,\theta_2,\phi_2)=C(\pi -\theta_1,\phi_1,\pi -\theta_2,\phi_2),
\end{eqnarray}
also under reflection of the coordinate system about x-y plane the
spherical harmonics transform as,
\begin{equation}
Y_{lm}(\pi-\theta,\phi)=(-1)^{l+m}Y_{lm}(\theta,\phi)\; .
\end{equation}
Therefore,  reflection symmetry demands,
\begin{equation}
P^{m_1}_{l_1}(\cos \theta_1) = P^{m_1}_{l_1}(\cos (\pi
-\theta_1))=P^{m_1}_{l_1}(-\cos \theta_1) .
\end{equation}
This implies $l_1 + m_1$ is even and similarly $l_2+m_2$. Here we
have used the symmetry property of Legendre polynomials,
$P_{lm}(-x)=(-1)^{l+m}P_{lm}(x)$. Interestingly, from symmetries of
spherical harmonics, one can show that n-fold symmetries are ruled
out for odd $n$ (see Appendix \ref{app:ruleout}).

Topologically compact universes exhibits even fold symmetry, but the
emergence of this fact from the symmetry of two-point correlation
pattern itself is nevertheless instructive. Therefore, we need to
look at the cases when \textit{n} is even.

\subsection{Even-fold Cylindrical symmetry}

Even fold symmetry refers to the case when $n$ is even. For compact
topologies this is always the case, for instance, Dirichlet domain
(DD) of a $T^{2}$ toroidal universe \cite{starobinsky92}, and a
$T^{3}$ have a 4-fold symmetry, that of a hexagonal prism has a
6-fold symmetry and a squeezed torus has 2 fold symmetry. This
symmetry puts another restriction on correlation function,
\begin{eqnarray}
C(\theta_1,\phi_1,\theta_2,\phi_2)=C(\theta_1,-\phi_1,\theta_2,-\phi_2) .
\end{eqnarray}
Hence most general correlation function under even-fold symmetry
is (see Appendix \ref{app:nfold}),

\begin{eqnarray*}
C^{(A)}(\theta_1,\phi_1,\theta_2,\phi_2) =
\sum_{m_1,m_2}f_{m_1,m_2}(\theta_1,\theta_2)
\delta_{m_1+m_2,nk}\nonumber\\ \cos(m_1\phi_1+m_2\phi_2) .
\end{eqnarray*}
Therefore Bipolar spherical harmonic coefficients are,
\begin{widetext}
\begin{eqnarray}\label{eq:evenbiposh}
A^{LM}_{l_1 l_2}=[1+(-1)^{l_1+l_2-L}]\sum_{m_1 m_2}
\sqrt{\frac{(2l_1+1)(2l_2+1)(l_1-m_1)!(l_2-m_2)!}{(l_1+m_1)!(l_2+m_2)!}}C^{LM}_{l_1
m_1 l_2
m_2}\delta_{m_1+m_2,nk}\nonumber\\\int^{2\pi}_{0}\int^{2\pi}_{0}
\int^{\pi}_{0}\int^{\pi}_{0}C(\theta_1,\phi_1,\theta_2,\phi_2) \cos
(m_1\phi_1+m_2\phi_2)P^{m_1}_{l_1}(\cos\theta_1)P^{m_2}_{l_2}
(\cos\theta_2)d{\phi_1}d{\phi_2}d{(\cos\theta_1)}d{(\cos\theta_2)}.
\end{eqnarray}
\end{widetext}
Reflection symmetry allows $A^{LM}_{l_1 l_2}$ only for even values
of $L$. For odd indices coefficients vanish
\begin{eqnarray}
A^{LM}_{l_1 l_2}=A^{LM}_{l_1 l_2}\delta_{M nk}\delta_{L 2a}, \ k =
0,1,2..., \ a=0,1,2,3...
\end{eqnarray}
Here $nk$ is even, therefore for an even-fold symmetry i.e.,
$(n=2,4,6,8...)$ $M$ is even and multipole moment $L\geq2$ picks up
only even values. Example: for a 2 fold symmetry possible values of
$M=2,4,6,8....$, similarly for 4-fold symmetry $M=4,8,12,16...$ and
so on. Another symmetry property of BipoSH is Eq. (\ref{eq:BipSym}),
so under even-fold rotational symmetry $l_1+l_2+m_1+m_2 =$ even
which implies $l_1+l_2$ is even hence $A^{LM}_{l_1 l_2}=A^{LM}_{l_2
l_1}$ i.e., BipoSH are symmetric under the exchange of $l_1$ and
$l_2$. Note that for all possible compact flat spaces BipoSH vanish
for odd indices, and the fact that two-point correlation function is
invariant under reflection about xy-plane plays a pivotal role in
restricting non-zero BipoSH only to even $L$'s. The rBiposh for even
universe with even fold symmetry are,
\begin{eqnarray}
A_{LM}=A_{LM}\delta_{M nk}\delta_{L 2a}.
\end{eqnarray}
Bipolar map will be,
\begin{eqnarray}
\Theta(\hat n)=\sum_{LM}A_{LM}Y_{LM}(\hat n)\delta_{M nk}\delta_{L
2a},
\end{eqnarray}
$L$ will take only even values and $M$ will run from $-L$ to $L$,
subsequently picking up even values.

\section{Bipolar Map:  Example of the  Bianchi template}\label{bianchi}

Now we will consider a Bianchi template as an example to show how a
Bipolar map looks like for a given temperature map. The choice of
Friedmann-Robertson-Walker (FRW) model as a model of our universe
was initially due to its simplicity, and later because of
observational evidence which strongly suggests universe to be
homogeneous and isotropic at large scales. However, the presently
observed isotropy may not necessarily hold in the past and the
universe may have been anisotropic in its early stages and tends to
FRW only later as it evolves. Bianchi models are the simplest
examples which have the property to isotropise as they evolve in
future. Bianchi classification contains $10$ equivalent classes
giving generic description of a homogeneous and anisotropic
cosmology~\cite{Barrow}. The most general Bianchi type which admits
FRW at late time are $VII_h$ and $IX$. However, the type $IX$
re-collapses after a finite time hence do not come arbitrarily close
to isotropy. Spiral pattern are characteristic signatures of $VII_0$
and $VII_h$ models~\cite{Barrow,Collins,Doroshkevich}. Jaffe et. al.
proposed Bianchi $VII_h$ models as an explanation of WMAP anomalies.
Since class $VII_h$ models resembles a universe with vorticity and
hence can lead to bounds on the universal rotation in cosmological
(CMB) data~\cite{Jaffe}. They proposed correction for some anomalies
in the first year maps from WMAP, however, introducing such
corrections induces other features like preferred direction and
violation of SI. Pontzen et al. calculated various temperature and
polarisation anisotropy patterns which may be formed in Bianchi
cosmologies \cite{Pontzen}. Ghosh et. al. analyzed the temperature
map for Bianchi $VII_h$ template ~\cite{Tuhin}. Given the
temperature map for Bianchi $VII_h$ template, here we see how a
Bipolar map actually looks like.

The temperature map for Bianchi $VII_h$ template is of the form
\begin{eqnarray}
\Delta T^B (\theta,\phi) =f_1(\theta)\sin\phi+f_2(\theta)\cos\phi ,
\end{eqnarray}
where super-script $B$ signifies Bianchi, and $f_1(\theta)$ and
$f_2(\theta)$ are parameters of the model which should be calculated
numerically~\cite{Barrow}.
\begin{widetext}
BipoSH for Bianchi template are,
\begin{eqnarray}
A^{LM}_{l_1 l_2}=\int^{\pi}_{0}\int^{\pi}_{0}\{\textit{W}_{l_1 l_2}(\theta_1,\theta_2)C^{LM}_{l_1 -1 l_2 1}\delta_{M0}+\textit{X}_{l_1 l_2}(\theta_1,\theta_2)C^{LM}_{l_1 -1 l_2 -1}\delta_{M-2}+\nonumber\\ \textit{Y}_{l_1 l_2}(\theta_1,\theta_2)C^{LM}_{l_1 1 l_2 -1}\delta_{M0}+\textit{Z}_{l_1 l_2}(\theta_1,\theta_2)C^{LM}_{l_1 1 l_2 1}\delta_{M2}\} d(\cos\theta_1)d(\cos\theta_2).
\end{eqnarray}
Therefore, rBiposh are,
\begin{eqnarray}
A_{LM}=\sum_{l_1 l_2}\int^{\pi}_{0}\int^{\pi}_{0}\{\textit{W}_{l_1 l_2}(\theta_1,\theta_2)C^{LM}_{l_1 -1 l_2 1}\delta_{M0}+\textit{X}_{l_1 l_2}(\theta_1,\theta_2)C^{LM}_{l_1 -1 l_2 -1}\delta_{M-2}+\nonumber\\ \textit{Y}_{l_1 l_2}(\theta_1,\theta_2)C^{LM}_{l_1 1 l_2 -1}\delta_{M0}+\textit{Z}_{l_1 l_2}(\theta_1,\theta_2)C^{LM}_{l_1 1 l_2 1}\delta_{M2}\} d(\cos\theta_1)d(\cos\theta_2),
\end{eqnarray}
where
\begin{eqnarray}
\textit{W}_{l_1 l_2}=\pi^{2}\sqrt{\frac{(2l_1+1)(2l_2+1)(l_1+1)!(l_2-1)!}{(4\pi)^{2}(l_1-1)!(l_2+1)!}}\{f_1(\theta_1)f_1(\theta_2)+i(f_1(\theta_1)f_2(\theta_2)-f_2(\theta_1)f_1(\theta_2))+\nonumber\\ f_2(\theta_1)f_2(\theta_2)\}P^{-1}_{l_1}(\cos\theta_1)P^{1}_{l_1}(\cos\theta_2)\nonumber,
\end{eqnarray}
\begin{eqnarray}
\textit{X}_{l_1 l_2}=\pi^{2}\sqrt{\frac{(2l_1+1)(2l_2+1)(l_1+1)!(l_2+1)!}{(4\pi)^{2}(l_1-1)!(l_2-1)!}}\{-f_1(\theta_1)f_1(\theta_2)+i(f_1(\theta_1)f_2(\theta_2)+f_2(\theta_1)f_1(\theta_2))+\nonumber\\ f_2(\theta_1)f_2(\theta_2)\}P^{-1}_{l_1}(\cos\theta_1)P^{-1}_{l_1}(\cos\theta_2)\nonumber,
\end{eqnarray}
\begin{eqnarray}
\textit{Y}_{l_1 l_2}=\pi^{2}\sqrt{\frac{(2l_1+1)(2l_2+1)(l_1-1)!(l_2+1)!}{(4\pi)^{2}(l_1+1)!(l_2-1)!}}\{f_1(\theta_1)f_1(\theta_2)+i(-f_1(\theta_1)f_2(\theta_2)+f_2(\theta_1)f_1(\theta_2))+\nonumber\\ f_2(\theta_1)f_2(\theta_2)\}P^{1}_{l_1}(\cos\theta_1)P^{-1}_{l_1}(\cos\theta_2)\nonumber,
\end{eqnarray}
\begin{eqnarray}
\textit{Z}_{l_1 l_2}=\pi^{2}\sqrt{\frac{(2l_1+1)(2l_2+1)(l_1-1)!(l_2-1)!}{(4\pi)^{2}(l_1+1)!(l_2+1)!}}\{-f_1(\theta_1)f_1(\theta_2)-i(f_1(\theta_1)f_2(\theta_2)+f_2(\theta_1)f_1(\theta_2))+\nonumber\\ f_2(\theta_1)f_2(\theta_2)\}P^{1}_{l_1}(\cos\theta_1)P^{1}_{l_1}(\cos\theta_2)\nonumber.
\end{eqnarray}
\end{widetext}
Hence,
\begin{equation}
A_{LM}=A_{LM}\delta_{M,k}\qquad k=0,\pm 2 \; .
\end{equation}

\begin{figure}[h]
\begin{center}
\mbox{\epsfig{figure=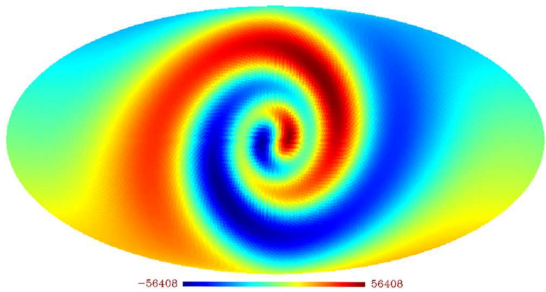,width=4.cm,angle=-90}} \caption{ {\it
Temperature map for Bianchi VII$_h$.  }} \label{temperaturemap}
\end{center}
\end{figure}
Possible values of $M$ are $0, \pm 2$. Note that $A_{LM}$ exists
only for $l_1 + l_2 - L=even$, and vanishes otherwise. Keeping in
mind the reality of two-point correlation function, i.e.,
$A_{LM}=(-1)^{M}A^{*}_{L -M}$, here we have $A_{L2}=A^{*}_{L -2}$.
Since $A_{LM}$ coefficients are complex numbers we can define,
$X_{LM}=\Re(A_{LM})$ and $Z_{LM}=\Im(A_{LM})$. Therefore, Bipolar
map for a Bianchi template looks like (see appendix
\ref{app:bianchi_temp}),
\begin{eqnarray}\nonumber
\Theta(\hat n)&=&\sum_{L}A_{L0}Y_{L0}(\theta,\phi)+2\sum_{L}X_{L
-2}G_{L}(\theta)\cos2\phi\\& -&
2\sum_{L}Z_{L-2}G_{L}(\theta)\sin2\phi,
\end{eqnarray}
where
\begin{eqnarray}
&& G_{L}(\theta) = \frac{1}{(\sin\theta)^{2}}
\sqrt{\frac{(L-1)L(L+1)(L+2)}{4\pi(2L+1)}} \\ \nonumber
&&\Big[\frac{P_{L-2}(\cos\theta)}{2L-1}-
\frac{2(2L+1)P_{L}(\cos\theta)}{(2L-1)(2L+3)}+\frac{P_{L+2}(\cos\theta)}{2L+3}\Big].
\end{eqnarray}
\begin{figure}[h]
\begin{center}
\mbox{\epsfig{figure=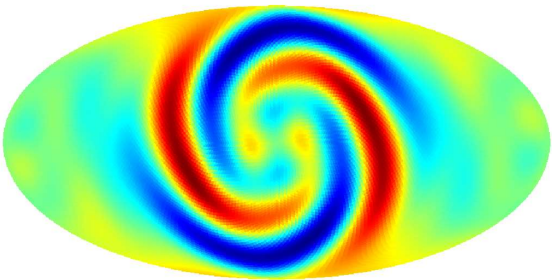,width=4.cm,angle=90}} \caption{ {\it
Bipolar map for Bianchi VII$_h$.  }} \label{Bipolarmap}
\end{center}
\end{figure}

Thus, a spiral pattern in temperature map will show up as double
spiral pattern in Bipolar map.

\section{Conclusion and discussion}
\label{disc}

Representation of correlation function of CMB anisotropy in terms of
Bipolar spherical harmonics provides a novel approach to study
violations of SI. Very recently the Bipolar representation has been
used to quantify anomalies in the analysis of WMAP seven - year data
\cite{wmap7}. These anisotropies can arise due to departure from FRW
metric (eg. Bianchi models), non-trivial spatial topologies (compact
spaces) or from primordial magnetic fields, among others. Here we
have studied various measurable quantities of Bipolar formalism to
quantify breakdown of SI.

We studied anisotropic homogeneous cosmologies which leave a
characteristic pattern on CMB. Like Bianchi $VII_h$ temperature map
which has a spiral pattern of a pair of cold and hot spots with a
dipole in azimuthal space. We found that the corresponding pattern in
Bipolar space becomes a double spiral having a quadrupole in azimuthal
space of the Bipolar map.

Another application is in case of homogeneous isotropic models where
an anisotropic topological identification has been imposed. As an
example, if the space is compact in one (or more) direction(s), the
statistical isotropy is broken due to introduction of preferred
direction(s).

We calculate BipoSH when this preferred direction is introduced. We
have shown here that for compact topologies, symmetry requirements can
restrict BipoSH to even multipole moments, i.e., BipoSH vanish for odd
indices for all kind of physically plausible models of flat
multi-connected universe.  Hyperbolic manifolds do not have the
desired symmetry and hence we expect odd multipoles to be non-zero in
these manifolds. Hence, we have a tool to distinguish different
topologies.  In case of homogeneous magnetic fields we have shown that
BipoSH's are restricted to even $L$ and $M=0$.

A new representation of Bipolar map has been proposed. Further work
needs to be done in this direction to extract new information from
this representation.

This technique can be applied to polarization maps and it may prove to
be a powerful method to decipher the topology of the universe,
something on which general relativity is completely silent.  The
Bipolar formalism can also be applied to various anisotropic universes
and can be used as a tool to distinguish various types of SI
breakdown.

\acknowledgements

The authors deeply regret the untimely demise of their colleague,
Himan Mukhopadhyay, and acknowledge her valuable contribution in this
work that originates and builds upon the excellent work carried out
during a graduate school project at IUCAA~\cite{himan}.  We also thank
Tuhin Ghosh for discussions at several stages of this work. SJ and NJ
would like to thank IUCAA for its hospitality and computational
facilities.

\appendix

\section{A review of topologically compact spaces}\label{app:review}

Topologically compact spaces break the statistical isotropy, thereby
introducing signatures in CMB correlation patterns. A compact
cosmological model, ${\cal M}$, is a \textit{Quotient space},
constructed by identifying points of standard FRW space under the
action of suitable discrete subgroup of motions $\varGamma$, of the
full isometry group \textit{G} of the FRW space. The isometry group
\textit{G} is the group of motions which preserves the distance
between points. The simply connected infinite FRW spatial hypersurface
with same constant curvature geometry is the universal cover (UC),
${\cal M}^u$, tiled by the copies of the compact space, ${\cal M}$. It
can be spherical (${\cal S}^3$), Euclidean (${\cal E}^3$) or
hyperbolic (${\cal H}^3$). The compact space for a given location of
observer is represented as \textit{Dirichlet domain} (DD), with the
observer at its basepoint. Any point \textbf{x} of the compact space
has an image, \textbf{$x_i$}= $\gamma_i$\textbf{x}, in each copy of DD
on the universal cover, where $\gamma_i\ \varepsilon\ \varGamma$. By
construction DD represents the compact space as a \textit{convex
polyhedron} with even number of faces identified pairwise under
$\varGamma$. In cosmology, DD around the observer represents the
universe as seen by the observer and the symmetries of the correlation
function are nothing but the symmetries of the corresponding DD.

\begin{figure}[ht]
\begin{center}
\mbox{\epsfig{figure=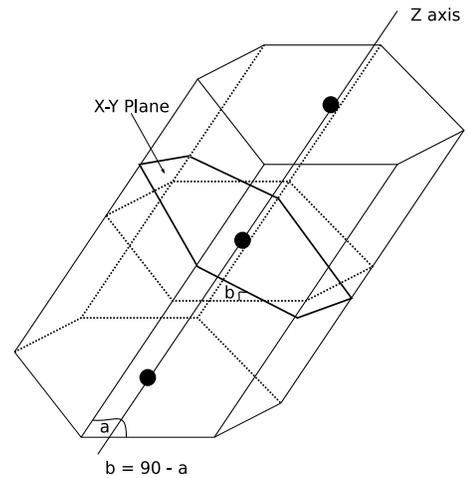,width=6.cm,angle=0}} \caption{
{\it Dirichlet domain of a squeezed torus. }}\label{fig:dirichlet}
\end{center}
\end{figure}

Correlation function of a scalar field, $\varPhi$, on a compact
manifold, ${\cal M}$, can be expressed as \citep{Chavel},
\begin{equation}
\xi^C_{\varPhi}(\textbf{x},\textbf{x}')=
\sum_{i}\sum_{j=1}^{m_i}P_{\varPhi}(k_i)\Psi_{ij}(\textbf{x})
\Psi^{*}_{ij}(\textbf{x}'),
\end{equation}
where
\begin{equation}
(\nabla^{2}+k^{2}_{i})\Psi_{ij}=0 .
\end{equation}
$\Psi_{i}$ are orthonormal set of eigenfunctions of Laplacian on the
hypersurface, having positive and discrete set of eigenvalues
$\{k^{2}_{i}\}$ $(k^{2}_o=0$ and $k^{2}_{i} < k^{2}_{i+1})$ with
multiplicities $m_i$. On a compact manifold, ${\cal M}$, the set of
eigenfunctions and eigenvalues are not always easy to obtain in closed
form (even numerically, for compact hyperbolic spaces). On the other
hand, eigenfunctions $\Psi^{u}_{j}(k,\bf{x})$ of the universal cover
(UC), ${\cal M}^u$, of a compact manifold usually known because of
their simplicity (e.g., ${\cal H}^3$, ${\cal S}^3$ and ${\cal E}^3$),
hence they can be used to compute the correlation functions
$\xi^C_{\varPhi}(\bf{x,x'})$ on the UC. For flat and hyperbolic UC's
the set of eigenvalues are continuous. The function $P_{\varPhi}(k_i)$
is the \textit{rms} amplitude of the eigenmode expansion of the field
$\varPhi$, whose information lies in the physical mechanism
responsible for the generation of $\varPhi$. The \textit{regularized
method of images}~\cite{Tarun}, describes how correlation function on
the compact manifold can be calculated once the correlation function
on the universal cover is known ~\citep{Tarun,Bond,Pogosyan}, which is
expressed as,
\begin{equation}
\xi^C_{\varPhi}(\bf{x,x'})=
\widetilde{\sum_{\gamma\varepsilon\varGamma}}\xi^{u}_{\varPhi}(\bf{x,x'})
.
\end{equation}
This implies that correlation function on a compact space, ${\cal M}$,
can be expressed as sum over the correlation function on its universal
covering space, ${\cal M}^u$, calculated between \textbf{x} and the
images $\gamma\bf{x'}(\gamma\ \varepsilon\ \varGamma)$ of
$\bf{x'}$. The local homogeneity and isotropy demands that the
correlation function on the UC is only a function of the distance
between two points $\bf{x}$ and $\bf{x'}$ i.e $\texttt{r}\equiv
\textit{d}(\textbf{x},\textbf{x}')$. The correlation function on a
compact universe with flat UC is,
\begin{eqnarray}
\xi^C_{\varPhi}(\textbf{x},\textbf{x}')=\sum_{i}
\int\frac{\it{dk}}{\it{k}}P_{\varPhi}(\it{k})\frac{\sin{\it{kd_i}}}{\it{kd_i}},
\end{eqnarray}
Here, $P_{\varPhi}(\it{k})$ can be determined from the early universe
physical mechanism and $\it{d_i}$ is the distance between the images
of $\bf{x}$ and $\bf{x'}$($\it{d_0}$ is the distance between original
points). Summation implies summing over all images. Hence, correlation
function depends not only on the distance between two points and the
distance of their images but symmetry defines both the pair to have
identical distance from their images i.e., take any two points on the
last scattering surface and their corresponding images about xy-plane,
correlation function will turn out to be invariant under this
reflection.
\begin{figure}[ht]
\begin{center}
\mbox{\epsfig{figure=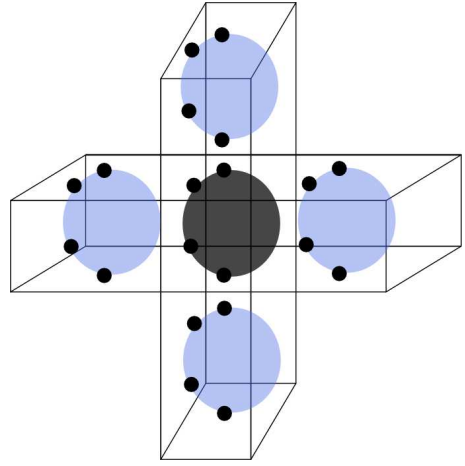,width=6.cm,angle=0}} \caption{{\it
Images of two-pairs of point which are mirror images of each other
about the XY plane in a $T^3$ space (one dimension suppressed).
}}\label{fig:t3_space}
\end{center}
\end{figure}
Figure (\ref{fig:t3_space}) illustrates this point for a $T^{3}$ universe.
The DD of a squeezed torus is shown in figure
(\ref{fig:dirichlet}). The choice of the axes here is a little bit
more non-trivial. The xy-plane is not parallel to any of the faces
of the DD or the FP, but still it would cut the LSS into two halves
in such a way that there will be symmetry under reflection about the
xy-plane and on the xy-plane there will be 2-fold rotational
symmetry.
However, here we point out that reflection symmetry does
not hold good for the compact spaces for which the opposite faces
are glued together with a twist~\cite{himan}.

Topology of the universe leaves characteristic signatures on CMB. If
the universe is finite and smaller than the distance to the last
scattering surface(LSS), then the signature of the topology of the
universe is imprinted on the CMB. For such a small universe LSS can
wrap around the universe and will self-intersect. The intersection
of the LSS, which is a 2-sphere with itself is a circle that will
appear twice in the cosmic microwave background. Hence, there might
exist pairs of circles which share correlated patterns of
temperature fluctuations. This \textit{circles in the
sky}~\cite{Cornish} method is a powerful and direct probe for
detecting non-trivial spatial topology. The correlated patterns
would be matching perfectly if the temperature fluctuation did not
depend on the direction of observation and if the patterns were not
distorted. However, the observed temperature fluctuations has
direction dependent components, i.e. the \textit{Doppler effect} and
the \textit{integrated Sachs-Wolfe effect}. Also observationally,
galaxy cut and foreground removals can also distort the matching.
However, one can search for such patterns in CMB correlation
function statistically. In a multi-connected space, there exist
preferred direction(s) so that global isotropy is broken. The
angular correlation will then depend on two directions of
observations and can also depend on the position of the observer.
This induces correlations between $a_{lm}$'s of different $l$ and
$m$. Thus, another indirect probe is to search such patterns or
signatures in the statistics of CMB temperature fluctuations.

\section{Bipolar map representation in terms of tripolar
spherical harmonics} \label{app:Bipolar_map}\label{app:rotation}

Bipolar map is defined as
\begin{equation}
\Theta(\hat n)=\sum_{LM}A_{LM}Y_{LM}(\hat n),
\end{equation}
where $A_{LM}=\sum_{l_1 l_2}A^{LM}_{l_1 l_2}$, therefore
\begin{equation}
\Theta(\hat n)=\sum_{LM}\sum_{l_1 l_2}A^{LM}_{l_1 l_2}Y_{LM}(\hat n)
\end{equation}
Now using the expansion of $A^{LM}_{l_1 l_2}$  we get
\begin{eqnarray*}
\Theta(\hat n)&=&\sum_{LM}\sum_{l_1 l_2}\int d\Omega_{\hat n_1}\int
d\Omega_{\hat n_2} C(\hat n_1,\hat n_2) \times \\ &&\{Y_{l_1}(\hat
n_1)\otimes Y_{l_2}(\hat n_2)\}^{*}_{LM}Y_{LM}(\hat n)
\end{eqnarray*}
tripolar scalar spherical harmonics are defined as
\begin{eqnarray}
&&\{Y_{l_1}(\hat n_1)\otimes\{Y_{L}(\hat n)\otimes Y_{l_2}(\hat
n_2)\}_{\lambda}\}_{00}=(-1)^{l_1+l_2+L}  \\ \nonumber && \delta_{\lambda
l_1}\sum_{m_1 m_2 M}\begin{pmatrix}
l_1 & L & l_2 \\
m_1 & M & m_2 \\
\end{pmatrix}Y_{l_1  m_1}(\hat n_1)Y_{L  M}(\hat n)Y_{l_2  m_2}(\hat n_2)
\end{eqnarray}
where $\begin{pmatrix}
l_1 & L & l_2 \\
m_1 & M & m_2 \\
\end{pmatrix}$
are Wigner-3j symbols  and are related to Clebsch Gordan
coefficients in the following way,
\begin{eqnarray*}
\CC_{l_{1}m_{1}l_{2}m_{2}}^{l_3
m_3}=(-1)^{l_1-l_2+m_3}\sqrt{2l_3+1}\begin{pmatrix}
l_1 & l_2 & l_3 \\
m_1 & m_2 & -m_3 \\
\end{pmatrix}
\end{eqnarray*}
Hence Bipolar map can be represented in terms of tripolar scalar
spherical harmonics,
\begin{eqnarray}
\Theta(\hat n)&=&\sum_{L, l_1, l_2}\int d\Omega_{\hat
n_1}d\Omega_{\hat n_2}C(\hat n_1,\hat n_2)(-1)^{l_1+l_2}\sqrt{2L+1}
\nonumber \\ && \delta_{\lambda L}\{Y_{l_1}(\hat n_1)\otimes
\{Y_{L}(\hat n)\otimes Y_{l_2}(\hat n_2)\}_{\lambda}\}_{00}
\end{eqnarray}
The representation of the Bipolar map in terms of tripolar harmonic
function makes the transformation properties of the bipolar map under
rotations explicit.

In a rotated sky map,
\begin{eqnarray}
\Theta'(\hat n)=\sum_{LM l_1 l_2}\int
 [ \sum_{l'_1,l'_2,L',M'}A^{L'M'}_{l'_1
l'_2}\sum_{M''}D^{L'}_{M''M'}(R) \nonumber \\  \{Y_{l'_1}(\hat n_1)\otimes
Y_{l'_2}(\hat n_2)\}_{L'M''} ] \times\nonumber\\ \{Y_{l_1}(\hat
n_1)\otimes Y_{l_2}(\hat n_2)\}^{*}_{LM}Y_{LM}(\hat n)d\Omega_{\hat
n_1}d\Omega_{\hat n_2}
\end{eqnarray}
Using orthogonality of Bipolar spherical harmonics
\begin{eqnarray}
\Theta'(\hat n)=\sum_{LM l_1
l_2}\sum_{l'_1,l'_2,L',M'}A^{L'M'}_{l'_1
l'_2}\sum_{M''}D^{L'}_{M''M'}(R) \nonumber \\ Y_{LM}\delta_{l_1
l'_1}\delta_{l_2 l'_2}\delta_{LL'}\delta_{MM'}
\end{eqnarray}

\begin{eqnarray}
\Theta'(\hat n)=\sum_{L M'}A_{L M'}\sum_{M}D^{L}_{M
M'}(R)Y_{LM}(\hat n)=\Theta(R\hat n)
\end{eqnarray}
Thus when correlation pattern is rotated by $``R"$, Bipolar map also
rotates by $``R"$.

\section{Cosmic Variance of Bipolar Quantities}\label{app:variance}
Cosmic variance is defined as the variance of estimator of an
observable constructed from a single sky map. In particular for
BipoSH
\begin{eqnarray}
\sigma^{2}({\tilde A^{LM}_{l_1 l_2}})=<({\tilde A^{LM}_{l_1
l_2}})^{2}> - {<{\tilde A^{LM}_{l_1 l_2}}>}^{2}.
\end{eqnarray}
Using Gaussianity of $\Delta T$, one can analytically compute the
variance of ${\tilde A^{LM}_{l_1 l_2}}$.
\begin{eqnarray}
{\tilde A^{LM}_{l_1 l_2}}=\sum_{m_1 m_2} a_{l_1 m_1} a^{*}_{l_2
m_2}(-1)^{m_2}C^{LM}_{l_1 m_1 l_2 -m_2}
\end{eqnarray}
therefore
\begin{eqnarray}
<\tilde A^{LM}_{l_1 l_2}\tilde A^{* LM}_{l_1
l_2}>&=&\sum_{m_1m_2}\sum_{m'_1 m'_2}<a_{l_1 m_1}a^{*}_{l_2
m_2}a^{*}_{l_1 m'_1}a_{l_2 m'_2}> \nonumber \\
&&(-1)^{m_2+m'_2}C^{LM}_{l_1 m_1 l_2 -m_2}C^{LM}_{l_1 m'_1 l_2
-m'_2}
\end{eqnarray}
Considering temperature field to be a Gaussian random field, one can expand the four-point correlation function in terms of two-point correlation function. Further, considering the fact that under statistical isotropy the covariance matrix is diagonal eq.(\ref{eq:SI}) the above equation reduces to
\begin{eqnarray}
<\tilde A^{LM}_{l_1 l_2}\tilde A^{* LM}_{l_1 l_2}>&=& C_{l_1}
C_{l_2} \delta_{l_1 l_2}(2l_1 +1) \delta_{L0} \delta_{M0} \nonumber
\\ &+& C_{l_1}C_{l_2} [1+(-1)^L \delta_{l_1 l_2}]
\end{eqnarray}
Also, we have
\begin{equation}
<{\tilde A^{LM}_{l_1 l_2}}> = (2 l_1+1)^{1/2} C_{l_1} \delta_{l_1
l_2} \delta_{L0} \delta_{M0}
\end{equation}
Hence the cosmic variance is
\begin{equation}
\sigma^{2}_{SI}({\tilde A^{LM}_{l_1
l_2}})=C_{l_1}C_{l_2}[1+(-1)^{L}\delta_{l_1 l_2}]
\end{equation}
Similarly for rBipoSH,
\begin{equation}
\sigma^{2}_{SI}(\tilde A_{LM})=\sum_{l_1 l_2}C_{l_1}C_{l_2}[1+(-1)^{l_1+l_2-L}]
\end{equation}

\section{Correlation function for Cylindrical symmetry}\label{app:zerofold}

Expansion of correlation function in terms of Bipolar spherical
harmonics is,
\begin{eqnarray}
C^{(A)}(\hat{n}_{1,}\hat{n}_{2}) = \sum_{l_{1},l_{2},L,M}
A_{l_{1}l_{2}}^{L M}\sum_{m_{1}m_{2}}\CC_{l_{1}m_{1}l_{2}m_{2}}^{LM}
\nonumber \\ Y_{l_{1}m_{1}}(\hat{n}_{1})\;
Y_{l_{2}m_{2}}(\hat{n}_{2})
\end{eqnarray}
Now rotational symmetry about z-axis for any arbitrary $\Delta \phi$
implies,
\begin{eqnarray}
C^{(A)}(\theta_1, \phi_1, \theta_2 ,\phi_2)=C^{A}(\theta_1
,{\phi_1}+\Delta\phi, \theta_2, {\phi_2}+\Delta\phi)
\end{eqnarray}
Therefore
\begin{eqnarray*}
\sum_{l_1,l_2,L,M,m_1,m_2}A_{l_{1}l_{2}}^{L M}\CC_{l_{1}m_{1}l_{2}m_{2}}^{LM}Y_{l_{1}m_{1}}(\theta_1,\phi_1)\;
Y_{l_{2}m_{2}}(\theta_2,\phi_2) \\
=
\sum_{l_1,l_2,L,M,m_1,m_2}A_{l_{1}l_{2}}^{L M}
\CC_{l_{1}m_{1}l_{2}m_{2}}^{LM}Y_{l_{1}m_{1}}(\theta_1,\phi_1+\Delta\phi) \\
Y_{l_{2}m_{2}}(\theta_2,\phi_2+\Delta\phi)
\end{eqnarray*}
which means
\begin{eqnarray}
e^{i(m_1+m_2)\Delta\phi}=1
\end{eqnarray}
therefore
\begin{eqnarray}
m_1+m_2=\frac{2k\pi}{\Delta\phi}, \ \ \ \ \ \ \ k=0,\pm1,\pm2....
\end{eqnarray}
for zero fold symmetry $m_1+m_2=0$ which means $m_1=-m_2$,
hence
\begin{eqnarray}
C^{(A)}(\hat{n}_{1,}\hat{n}_{2})=\sum_{l_{1},l_{2},L,M,m_1,m_2} A_{l_{1}l_{2}}^{L M}\CC_{l_{1}m_{1}l_{2}m_{2}}^{LM}Y_{l_{1}m_{1}}(\hat{n}_{1}) \nonumber \\
Y_{l_{2}m_{2}}(\hat{n}_{2})\delta_{m_1,-m_2}
\end{eqnarray}
using the expansion of spherical harmonics in terms of associated Legendre polynomials,
\begin{eqnarray}
Y_{l m}(\theta,\phi)=e^{im\phi}\sqrt{\frac{(2l+1)(l-m)!}{4\pi (l+m)!}}P^{m}_l(\cos\theta)
\end{eqnarray}
therefore correlation function will be,
\begin{eqnarray}
C^{(A)}(\hat{n}_{1,}\hat{n}_{2})=\sum_{m}f_{m}(\theta_1,\theta_2)e^{im(\phi_1-\phi_2)}
\end{eqnarray}
where
\begin{eqnarray}
f_{m}(\theta_1,\theta_2)=\frac{1}{4\pi}\sum_{l_{1},l_{2},L}A_{l_{1}l_{2}}^{L
M}\CC_{l_{1}ml_{2} -m}^{L M}\delta_{M0} \nonumber \\
\sqrt{\frac{(2l_1+1)(2l_2+1)(l_1-m)!(l_2+m)!}{l_1+m)!(l_2-m)!}}
\nonumber \\ P^{m}_{l_1}(\cos\theta_1)P^{-m}_{l_2}(\cos\theta_2)
\end{eqnarray}
Symmetry ensures,
\begin{eqnarray}
C^{(A)}(\theta_1, \phi_1, \theta_2 ,\phi_2)=C^{A}(\theta_1, -\phi_1,
\theta_2 ,-\phi_2)
\end{eqnarray}
Imposing this symmetry we get,
\begin{eqnarray}\label{eq:cylcorr}
C^{(A)}(\theta_1, \phi_1, \theta_2
,\phi_2)=\sum_{m}f_{m}(\theta_1,\theta_2)\cos m(\phi_1-\phi_2)
\end{eqnarray}
This is the most general correlation function under zero fold rotational symmetry.
Using
\begin{eqnarray}
\nonumber A^{L M}_{l_1 l_2} = \int d\Omega_{\hat{n}_1}d\Omega_{\hat{
n}_2}C (\hat{n}_1, \hat{n}_2)\{ Y_{l_1} (\hat{n}_1)
\otimes Y_{l_2} (\hat{n}_2)\}^{*}_{LM}
\end{eqnarray}
and eq.(\ref{eq:cylcorr}) we get,
\begin{widetext}
\begin{eqnarray}\label{eq:bipocoeff}
A^{L M}_{l_1 l_2} =[1+(-1)^{l_1+l_2-L}]\sum_{m}(-1)^{m}\frac{\sqrt{(2l_1+1)(2l_2+1)(l_1-m)!(l_2-m)!}}{(4\pi)^{2}(l_1+m)!(l_2+m)!}C^{LM}_{l_1 ml_2 -m}\delta_{M0}\nonumber\\
\int^{\pi}_{0}\int^{\pi}_{0}P_{l_1 m}(\cos\theta_1)P_{l_2
m}(\cos\theta_2)
f_{m}(\theta_1,\theta_2)d(\cos\theta_1)d(\cos\theta_2).
\end{eqnarray}
\end{widetext}

\section{ n-fold cylindrical symmetry}\label{app:nfold}

Correlation function in such a case is,
\begin{equation}
C^{(A)}(\theta_1,\phi_1,\theta_2,\phi_2)=
C^{(A)}(\theta_1,\phi_1+\frac{2\pi}{n},\theta_2,\phi_2+\frac{2\pi}{n})\,.
\end{equation}

%\begin{widetext}
This implies,
\begin{eqnarray*}
\sum_{l_1 l_2 m_1 m_2 L M} A^{LM}_{l_1 l_2} C^{LM}_{l_1 m_1 l_2 m_2}
Y_{l_1 m_1}(\theta_1,\phi_1) Y_{l_2 m_2}(\theta_2,\phi_2) = \qquad \qquad \\
\sum_{l_1 l_2 m_1 m_2 L M} A^{LM}_{l_1 l_2} C^{LM}_{l_1 m_1 l_2 m_2}
Y_{l_1 m_1}(\theta_1,\phi_1+\frac{2\pi}{n}) Y_{l_2
m_2}(\theta_2,\phi_2 + \frac{2\pi}{n})
\end{eqnarray*}
%\end{widetext}

Hence, $e^{i(m_1+m_2)\frac{2\pi}{n}}=1$, which implies $m_1+m_2 =
nk,  k=0,\pm 1,\pm 2,\pm 3.....$.

Most general form of correlation function will be,
\begin{eqnarray*}
C^{(A)}(\theta_1,\phi_1,\theta_2,\phi_2) = \qquad \qquad \qquad\qquad\qquad   \\
\sum_{m_1 m_2}f_{m_1
m_2}(\theta_1,\theta_2)e^{i(m_1\phi_1+m_2\phi_2)}\delta_{m_1+m_2,nk}
\end{eqnarray*}

\begin{widetext}
where
\[f_{m_1 m_2}(\theta_1,\theta_2)=\frac{1}{4\pi}\sum_{l_1l_2 L M}
\sqrt{\frac{(2l_1+1)(2l_2+1)(l_1-m_1)!(l_2-m_2)!}{(l_1+m_1)!(l_2+m_2)!}}A^{LM}_{l_1
l_2}C^{LM}_{l_1 m_1 l_2
m_2}P^{m_1}_{l_1}(\cos\theta_1)P^{m_2}_{l_2}(\cos\theta_2) \]
\end{widetext}
Demanding explicitly the two fold  symmetry that holds for all even-fold symmetry,
\begin{eqnarray}
C(\theta_1,\phi_1,\theta_2,\phi_2)=C(\theta_1,-\phi_1,\theta_2,-\phi_2)
\end{eqnarray}
this symmetry rules out the presence of sine terms in correlation
function. Hence for even-fold symmetry correlation function reduces
to,
\begin{eqnarray}
C^{(A)}(\theta_1,\phi_1,\theta_2,\phi_2)=\sum_{m_1,m_2}f_{m_1,m_2}(\theta_1,\theta_2)
\nonumber \\ \delta_{m_1+m_2,nk}\cos(m_1\phi_1+m_2\phi_2)
\end{eqnarray}

\section{Absence of odd-fold symmetries for compact spaces}\label{app:ruleout}
Let us take a compact space with reflection symmetry. It would
demand that
\begin{eqnarray}\label{eq:1}
C(\pi-\theta_1,\phi_1,\pi-\theta_2,\phi_2)=C(\theta_1,\phi_1,\theta_2,\phi_2) \,.
\end{eqnarray}
It can be shown that
\begin{eqnarray}\label{eq:2}
C(\theta_1,\phi_1,\theta_2,\phi_2) &&= \sum_{l_1 l_2 m_1 m_2 L
M}(-1)^{m_2}<a_{l_1 m_1}a^{*}_{l_2 m_2}> \times \nonumber \\ &&
C^{LM}_{l_1 m_1 l_2 -m_2}\{Y_{l_{1}}(\hat{n}_{1})\otimes
Y_{l_{2}}(\hat{n}_{2})\}_{LM} \,.
\end{eqnarray}
The symmetry of spherical harmonics would ensure that
\begin{eqnarray}
&&\{Y_{l_{1}}(\pi-\theta_1,\phi_1)\otimes
Y_{l_{2}}(\pi-\theta_2,\phi_2)\}_{LM} \nonumber
\\&&=(-1)^{m_1+m_2}\{Y_{l_{1}}(\theta_1,\phi_1)\otimes
Y_{l_{2}}(\theta_2,\phi_2)\}_{LM} \,.
\end{eqnarray}
This put together with equations (\ref{eq:1}) and (\ref{eq:2})
indicates that $m_1$ must be an even number. Since they are dummy
indices, $m_2$ would be even too. Now let us consider such a space
with n fold symmetry. Evidently
\begin{equation}\label{eq:3}
C(\theta_1,\phi_1+\frac{2\pi}{n},\theta_2,\phi_2+\frac{2\pi}{n}) =
C(\theta_1,\phi_1,\theta_2,\phi_2)
\end{equation}
since we know
\begin{eqnarray}
\{Y_{l_{1}}(\theta_1,\phi_1+\frac{2\pi}{n}) \otimes
Y_{l_{2}}(\theta_2,\phi_2+\frac{2\pi}{n})\}_{LM}= \nonumber \\ \exp
(\frac{2\pi(m_1+m_2)}{n})\{Y_{l_{1}}(\theta_1,\phi_1)\otimes
Y_{l_{2}}(\theta_2,\phi_2)\}_{LM} \,.
\end{eqnarray}
For Eq. (\ref{eq:3}) to hold $(m_1+m_2)/n$ must be even. Since $m_1$
and $m_2$ are even, n has to be even too.

\section{Bianchi template}\label{app:bianchi_temp}

The temperature map for Bianchi template is written as
\begin{eqnarray}
\Delta T (\theta,\phi) = f_1(\theta)\sin\phi+f_2(\theta)\cos\phi\,.
\end{eqnarray}
Bipolar map can be expressed as,
\begin{eqnarray*}
\Theta(\hat n)=\sum_{LM}\sum_{l_1 l_2}\sum_{m_1 m_2}\int
d\Omega_{\hat n_1} d\Omega_{\hat n_2}<\Delta T(\hat n_1)\Delta
T(\hat n_2)> \\ C^{LM}_{l_1 m_1 l_2 m_2}Y^{*}_{l_1 m_1}(\hat
n_1)Y^{*}_{l_2 m_2}(\hat n_2)Y_{LM}(\hat n)\,.
\end{eqnarray*}
The integrals over $\phi$ contribute only for $m=\pm1$ otherwise it
vanishes. Constraint on the values of $m_1=\pm1$ and $m_2=\pm1$,
admits only $M=0,\pm2$. Reduced Bipolar coefficient is then
\begin{eqnarray*}
A_{LM}=\sum_{l_1 l_2}\sum_{m_1 m_2}\int d\Omega_{\hat
n_1}d\Omega_{\hat n_2}<\Delta T(\hat n_1)\Delta T(\hat n_2)> \\
(-1)^{m_1 + m_2} C^{LM}_{l_1 m_1 l_2 m_2}Y_{l_1 -m_1}(\hat
n_1)Y_{l_2 -m_2}(\hat n_2)\,.
\end{eqnarray*}
$A_{LM}$ exists only for $l_1+l_2-L=even$, and vanishes otherwise
and the reality condition demands $A_{LM}$, i.e.,
$A_{LM}=(-1)^{M}A^{*}_{L -M}$. Now Bipolar map is
\begin{equation}
\Theta(\hat n)=\sum_{LM}A_{LM}Y_{LM}({\hat n})
\end{equation}
but for Bianchi template it will be
\begin{eqnarray*}
\Theta(\hat n)=\sum_{L}A_{L0}Y_{L0}(\hat
n)+\sum_{L\geq2}A_{L2}Y_{L2}(\hat
n)\\+\sum_{L\geq2}A_{L-2}Y_{L-2}(\hat n)
\end{eqnarray*}
which can be written as
\begin{eqnarray}\label{bitemp}
\Theta(\hat n)=\sum_{L}A_{L0}Y_{L0}(\hat
n)+\sum_{L\geq2}A^{*}_{L-2}Y^{*}_{L-2}(\hat n) \nonumber \\
+\sum_{L\geq2}A_{L-2}Y_{L-2}(\hat n)
\end{eqnarray}
Since $A_{LM}'s$ are complex numbers,  we define
\begin{equation}
A_{LM}=X_{LM}+iZ_{LM} \; \text{and} \ A^{*}_{LM}=X_{LM}-iZ_{LM}
\end{equation}
and the Bipolar map (\ref{bitemp}) can then be written as
\begin{eqnarray*}
\theta(\hat n)=\sum_{L}A_{L0}Y_{L0}(\hat n)&+&\sum_{L}X_{L
-2}\big(Y^{*}_{L -2}(\hat n)+Y_{L -2}(\hat n)\big) \nonumber \\
&+&i\sum_{L}Z_{L-2}\big(Y_{L -2}(\hat n)-Y^{*}_{L -2}(\hat n)\big) .
\end{eqnarray*}
Defining,
\begin{eqnarray*}
G_{L}(\theta) =
\frac{1}{(\sin\theta)^{2}}\sqrt{\frac{(L-1)L(L+1)(L+2)}{4\pi(2L+1)}}
\qquad \qquad \nonumber \\
\Big[\frac{P_{L-2}(\cos\theta)}{2L-1}-\frac{2(2L+1)
P_{L}(\cos\theta)}{(2L-1)(2L+3)}+\frac{P_{L+2}(\cos\theta)}{2L+3}\Big]
\end{eqnarray*}
the Bipolar map is represented as
\begin{eqnarray}
\Theta(\theta,\phi)=\sum_{L}A_{L0}Y_{L0}(\theta,\phi)+\sum_{L}X_{L
-2}G_{L}(\theta)2\cos2\phi \nonumber \\
-\sum_{L}Z_{L-2}G_{L}(\theta)2\sin2\phi
\end{eqnarray}
where we have used expansion of spherical harmonics in terms of
associated Legendre polynomials
\begin{eqnarray*}
Y_{lm}(\theta,\phi)=e^{im\phi}\sqrt{\frac{(2l+1)(l-m)!}{4\pi
(l+m)!}}P^{m}_{l}(\cos\theta)
\end{eqnarray*}
and
\begin{eqnarray*}
Y_{l \pm2}(\theta,\phi) =
\frac{e^{i\pm2\phi}}{(\sin\theta)^{2}}\sqrt{\frac{(L-1)L(L+1)(L+2)}{4\pi(2L+1)}}
\nonumber \\
\Big[\frac{P_{L-2}(\cos\theta)}{2L-1} -
\frac{2(2L+1)P_{L}(\cos\theta)}{(2L-1)(2L+3)}+\frac{P_{L+2}(\cos\theta)}{2L+3}\Big].
\end{eqnarray*}

\section{Useful Mathematical Relations}\label{app:math}

Orthonormality of spherical harmonics
\begin{eqnarray}
 \int d\Omega_{\hat{n}}\, Y_{l_1m_1}(\hat{n})Y_{l_2m_2}^*(\hat{n})&=&\delta_{l_1l_2}\delta_{m_1m_2}
\end{eqnarray}
Symmetry property of spherical harmonics
\begin{eqnarray}
Y_{lm}^*(\hat{n})=(-1)^mY_{l-m}(\hat{n}).
\end{eqnarray}
Spherical harmonic expansion of Legendre polynomials
\begin{eqnarray}
P_l(\hat{n}\cdot \hat{n}')=\frac{4\pi}{2l+1}\sum_{m=-l}^{l}{Y_{lm}^*(\hat{n})Y_{lm}( \hat{n}')}.
\end{eqnarray}
Property of legendre polynomial
\begin{eqnarray}
P^{-m}_{l}=(-1)^{m}\frac{(l-m)!}{(l+m)!}P^{m}_{l} \,.
\end{eqnarray}

Symmetry properties of Clebsch-Gordan coefficients
\begin{eqnarray}
C_{a \alpha b \beta}^{c \gamma} &=& (-1)^{a+b-c} C_{ b \beta a \alpha }^{c \gamma} \,,
\\ \nonumber
C_{a \alpha b \beta}^{c \gamma} &=& (-1)^{a+b-c} C_{a -\alpha b -\beta}^{c -\gamma} \,.
\end{eqnarray}
Summation rules of Clebsch-Gordan coefficients
\begin{eqnarray}\label{eq:trieq}
\sum_{\alpha \beta} C_{a \alpha b \beta}^{c \gamma} C_{a \alpha b
\beta}^{c' \gamma'} &=& \delta_{cc'} \delta_{\gamma \gamma'}
\{abc\}\{abc'\}
\nonumber \\
\sum_{a \gamma} C_{a \alpha b \beta}^{c \gamma} C_{a \alpha b' \beta'}^{c \gamma} &=& \frac{2c+1}{2b+1} \delta_{bb'} \delta_{\beta \beta'}\{abc\}\{ab'c\}
\nonumber \\
\sum_{c\gamma} C_{a \alpha b \beta}^{c \gamma} C_{a \alpha' b
\beta'}^{c \gamma} &=& \delta_{\alpha \alpha'}\delta_{\beta
\beta'}\{abc\}\,
\nonumber \\
\sum_{b}(-1)^{a-b}\CC^{c 0}_{a b a -b} &=& \prod_{a}\delta_{c 0}
\end{eqnarray}
where
\begin{equation}
\prod_{abc.....}=[(2a+1)(2b+1)....(2c+1)]^{1/2}
\end{equation}
and where $\{abc\}$ is defined by
\begin{equation}
\{abc\} = \left\{ \begin{array}{rl}
 1 & \mbox{ if $a+b+c$ is an integer} \\
 0 & \mbox{ otherwise}
       \end{array} \right.
\end{equation}
and where $a, b$ and $c$ satisfy triangle inequality $|a-b| \leq c
\leq (a+b)$.

Tripolar spherical harmonics are expressed as,
\begin{eqnarray*}
&&\{Y_{l_1}(\hat n_1)\otimes\{Y_{l_2}(\hat n_2)\otimes Y_{l_3}(\hat
n_3)\}_{l_{23}}\}_{LM}= \\\nonumber && \sum_{m_1 m_2 m_3 m_{23}}
C^{LM}_{l_1 m_1 l_{23} m_{23}}C^{l_{23}m_{23}}_{l_2 m_2 l_3
m_3}Y_{l_1 m_1}(\hat n_1)Y_{l_2 m_2}(\hat n_2) Y_{l_3 m_3}(\hat n_3)
\qquad \qquad .
\end{eqnarray*}
Tripolar scalar spherical harmonics are defined as
\begin{eqnarray}
&&\{Y_{l_1}(\hat n_1)\otimes\{Y_{L}(\hat n)\otimes Y_{l_2}(\hat
n_2)\}_{\lambda}\}_{00}=(-1)^{l_1+l_2+L}  \\ \nonumber && \delta_{\lambda
l_1}\sum_{m_1 m_2 M}\begin{pmatrix}
l_1 & L & l_2 \\
m_1 & M & m_2 \\
\end{pmatrix}Y_{l_1  m_1}(\hat n_1)Y_{L  M}(\hat n)Y_{l_2  m_2}(\hat n_2)\,.
\end{eqnarray}
where $\begin{pmatrix}
l_1 & L & l_2 \\
m_1 & M & m_2 \\
\end{pmatrix}$ are Wigner-3j symbols.
\newline Orthogonality of tripolar spherical harmonics, is given as
\begin{eqnarray*}
\int\int\int&& d\Omega_{\hat n_1}d\Omega_{\hat n_2}d\Omega_{\hat
n_3}\{Y_{l_1}(\hat n_1)\otimes\{Y_{l_2}(\hat n_2)\otimes
Y_{l_3}(\hat n_3)\}_{\lambda}\}_{LM}\\ && \{Y_{l'_1}(\hat
n_1)\otimes\{Y_{l'_2}(\hat n_2)\otimes Y_{l'_3}(\hat
n_3)\}_{\lambda'}\}^{*}_{L'M'} \\ &&= \delta_{l_1 l'_1}\delta_{l_2
l'_2}\delta_{l_3 l'_3}\delta_{\lambda \lambda'}\delta_{L
L'}\delta_{M M'}\,.
\end{eqnarray*}


\begin{thebibliography}{99}

\bibitem{preferreddirections}
  A.~de Oliveira-Costa, M.~Tegmark, M.~Zaldarriaga and A.~Hamilton,
  Phys.\ Rev.\  D {\bf 69}, 063516 (2004)
  [arXiv:astro-ph/0307282];
  C.~J.~Copi, D.~Huterer and G.~D.~Starkman,
  Phys.\ Rev.\  D {\bf 70}, 043515 (2004)
  [arXiv:astro-ph/0310511];
  D.~J.~Schwarz, G.~D.~Starkman, D.~Huterer and C.~J.~Copi,
  Phys.\ Rev.\ Lett.\  {\bf 93}, 221301 (2004)
  [arXiv:astro-ph/0403353];
  S.~Prunet, J.~P.~Uzan, F.~Bernardeau and T.~Brunier,
  Phys.\ Rev.\  D {\bf 71}, 083508 (2005)
  [arXiv:astro-ph/0406364];
  H.~K.~Eriksen, A.~J.~Banday, K.~M.~Gorski and P.~B.~Lilje,
  Astrophys.\ J.\  {\bf 622}, 58 (2005)
  [arXiv:astro-ph/0407271];
  K.~Land and J.~Magueijo,
  Phys.\ Rev.\ Lett.\  {\bf 95}, 071301 (2005)
  [arXiv:astro-ph/0502237];
  T.~R.~Jaffe, A.~J.~Banday, H.~K.~Eriksen, K.~M.~Gorski and F.~K.~Hansen,
  Astrophys.\ J.\  {\bf 629}, L1 (2005)
  [arXiv:astro-ph/0503213];
  C.~J.~Copi, D.~Huterer, D.~J.~Schwarz and G.~D.~Starkman,
  Mon.\ Not.\ Roy.\ Astron.\ Soc.\  {\bf 367}, 79 (2006)
  [arXiv:astro-ph/0508047];
  K.~Land and J.~Magueijo,
  Mon.\ Not.\ Roy.\ Astron.\ Soc.\  {\bf 367}, 1714 (2006)
  [arXiv:astro-ph/0509752];
  A.~Bernui, T.~Villela, C.~A.~Wuensche, R.~Leonardi and I.~Ferreira,
  Astron.\ Astrophys.\  {\bf 454}, 409 (2006)
  [arXiv:astro-ph/0601593];
  L.~R.~Abramo, A.~Bernui, I.~S.~Ferreira, T.~Villela and C.~A.~Wuensche,
  Phys.\ Rev.\  D {\bf 74}, 063506 (2006)
  [arXiv:astro-ph/0604346];
  J.~Magueijo and R.~D.~Sorkin,
  Mon.\ Not.\ Roy.\ Astron.\ Soc.\ Lett.\  {\bf 377}, L39 (2007)
  [arXiv:astro-ph/0604410];
  C.~G.~Park, C.~Park and J.~R.~I.~Gott,
  Astrophys.\ J.\  {\bf 660}, 959 (2007)
  [arXiv:astro-ph/0608129];
  D.~Huterer,
  New Astron.\ Rev.\  {\bf 50}, 868 (2006)
  [arXiv:astro-ph/0608318];
  P.~Vielva, Y.~Wiaux, E.~Martinez-Gonzalez and P.~Vandergheynst,
  New Astron.\ Rev.\  {\bf 50}, 880 (2006)
  [arXiv:astro-ph/0609147];
  K.~Land and J.~Magueijo,
  Mon.\ Not.\ Roy.\ Astron.\ Soc.\  {\bf 378}, 153 (2007)
  [arXiv:astro-ph/0611518];
  C.~Gordon, W.~Hu, D.~Huterer and T.~Crawford,
  Phys.\ Rev.\  D {\bf 72}, 103002 (2005)
  [arXiv:astro-ph/0509301];
  J.~G.~Cresswell, A.~R.~Liddle, P.~Mukherjee and A.~Riazuelo,
  Phys.\ Rev.\  D {\bf 73}, 041302 (2006)
  [arXiv:astro-ph/0512017].
  S.~H.~S.~Alexander,
  arXiv:hep-th/0601034.

\bibitem{Eriksen:2003db}
H.~K.~Eriksen, F.~K.~Hansen, A.~J.~Banday, K.~M.~Gorski and
P.~B.~Lilje, Astrophys.\ J.\  {\bf 605}, 14 (2004) [Erratum-ibid.\
{\bf 609}, 1198 (2004)] [arXiv:astro-ph/0307507].

\bibitem{Hansen:2004vq} F.~K.~Hansen, A.~J.~Banday and K.~M.~Gorski,
  arXiv:astro-ph/0404206.

\bibitem{G.Ellis} G. F. R. Ellis, Gen. Rel. Grav.{\bf 2}, 7 (1971).

\bibitem{Lachieze} M. Lachieze-Rey and J. P. Luminet, Phys. Rept. {\bf 254}, 135 (1995) [arXiv:gr-qc/9605010].

\bibitem{Gott} J. R. Gott, Mon. Not. R. Astr. Soc. {\bf 193}, 153 (1980).

\bibitem{Cornish1} N. J. Cornish, D. N. Spergel and G. D. Starkman, Phys. Rev. Lett. {\bf 77}, 215 (1996).

\bibitem{Levin} J. Levin, Phys. Rept. {\bf 365}, 251 (2002) [arXiv:gr-qc/0108043].

\bibitem{Linde} A. Linde, JCAP {\bf 0410}, 004 (2004) [arXiv:hep-th/0408164].

\bibitem{Souradeep} T. Souradeep, \emph{Spectroscopy of Cosmic Topology} [arXiv:gr-qc/0609026].

\bibitem{Ellis} G. Ellis and M. MacCallum, Commun. Math. Phys. {\bf
12}, 108 (1969).

\bibitem{Barrow} J. D. Barrow, R. Juszkiewicz and D. H. Sonoda, Mon.
Not. R. astr. Soc., {\bf 213} 917 (1985).

\bibitem{Tuhin} T. Ghosh, A. Hajian and T. Souradeep, Phys.\ Rev.\
D {\bf 75}, 083007 (2007).

\bibitem{Ratra} B. Ratra, ApJ {\bf 391}, L1 (1992).

\bibitem{Ruth} R. Durrer, T. Kahniashvili and A. Yates. Phys. Rev. D {\bf 58}, 123004 (1998).

\bibitem{Pogosyan} J. R. Bond, D. Pogosyan, T. Souradeep, Phys.\ Rev.\  D {\bf 62}, 043006 (2000).

\bibitem{AH-TS-03} A. Hajian and T. Souradeep, ApJ  {\bf 597} L5 (2003).

\bibitem{Amirthesis} A. Hajian, `Cosmology with CMB anisotropy', PhD
thesis (2006).

\bibitem{AH-TS-04} T. Souradeep A Hajian, Pramana {\bf 62}, 793-796 (2004).

\bibitem{AH-TS-NC} A Hajian, T. Souradeep and N. Cornish, ApJ {\bf 618}, L63-L66 (2004).

\bibitem{AH-TS-05} A Hajian and T. Souradeep [astro-ph/0501001].

\bibitem{SB-AH-TS} S. Basak, A. Hajian and T. Souradeep Phys.\ Rev.\  D {\bf 74}, 021301 (2006).

\bibitem{AH-TS-06} A. Hajian and T. Souradeep, Phys.\ Rev.\  D {\bf 74}, 123521 (2006).

\bibitem{varsha} D. A. Varshalovich, A. N. Moskalev and V. K. Khersonskii,
\emph{ Quantum Theory of Angular Momentum} (Singapore: World
Scientific, 1988).

\bibitem{Tarun} T. Souradeep, D. Pogosyan, J. R. Bond, Proc. of
XXXIIIrd Recontres de Moriond "Fundamental Parameters in Cosmology",
Jan. 17-24,1998, Les Arc, France [astro-ph/9804042].

\bibitem{Bond} J. R. Bond, D. Pogosyan, T. Souradeep, Phys.\ Rev.\  D {\bf 62}, 043005 (2000).

\bibitem{Weeks} J. Weeks, J. P. Luminet, A. Riazuelo and R. Lehoucq, Mon. Not. Roy. Astron. Soc. {\bf 352}, 258 (2004) [arXiv:astro-ph/0312312].

\bibitem{hogg} D. W. Hogg, D. J. Eisenstein, M. R. Blanton, N. A. Bachall, J. Brinkmann, J. E. Gunn and D. P. Schneider, Astrophys. J. {\bf 624}, 54 (2005).

\bibitem{Bennet03} C. L. Bennet et. al., ApJSS {\bf 148} 1, (2003).

\bibitem{spergel} D. N. Spergel et al. [WMAP collaboration] Astrophys. J. Suppl. {\bf 148}, 175 (2003).

\bibitem{krasinski} A. Krasinski, \emph{Inhomogeneous Cosmological
Models}, (Cambridge University Press, Cambridge, England, 1996).

\bibitem{Rubenstein} D. Grasso and H. R. Rubenstein, Phys. Rept. {\bf 348}, 163 (2001).

\bibitem{Ferreira:1997} P. G. Ferreira, and J. Magueijo.
Phys.\ Rev.\  D {\bf 56}, 4578  (1997).

\bibitem{starobinsky92} A. A. Starobinsky, JETP Lett.
{\bf 57}, 622 (1992).

\bibitem{Collins} C. B. Collins and S. W. Hawking, Mon. Not. R. astr. Soc. {\bf 162}, 307 (1973).

\bibitem{Doroshkevich}A. G. Doroshkevich, V. N. Lukash and I. D. Novikov, Soviet Astr. {\bf 18}, 554 (1975).

\bibitem{Jaffe} T. R. Jaffe, A. J. Banday, H. K. Eriksen, K. M. Gorski and F. K. Hansen, Astrophys. J. {\bf 629}, L1 (2005) [arXiv:astro-ph/0503213].

\bibitem{Pontzen} A. Pontzen, Phys.\ Rev.\  D {\bf 79}, 103518
(2009).

\bibitem{himan} H. Mukhopadhyay, `Cosmic spectroscopy of CMB correlation patterns',
\url{http://meghnad.iucaa.ernet.in/~tarun/papers/students/himan_project.pdf}

\bibitem{Chavel} I. Chavel, Eigenvalues in Riemannian geometry, (Academic Press, 1984).

\bibitem{Cornish} N. J. Cornish, D. N. Spergel and G. D. Starkman, Class. Quant. Grav. {\bf 15}, 2657 (1998) [arXiv:gr-qc/9602039].

\bibitem{wmap7} C. L. Bennett et al. [arxiv:1001.4758].


\end{thebibliography}
\end{document}